\documentclass[12pt]{article}

\usepackage{amsmath}

\title{Linear analysis and crossphase dynamics in the $\nabla T_e$-driven CTEM fluid model}
\author{M. Leconte$^1$, Lei Qi$^1$ and J. Anderson$^2$ \\
$^1$ Korea Institute of Fusion Energy (KFE), Daejeon 34133, South Korea \\
$^2$ Department of Space, Earth and Environment Sciences, \\ Chalmers University of Technology, SE-412 96 G${\rm \ddot o}$teborg, Sweden \\
\quad\\
E-mail: mleconte@kfe.re.kr}

\newcommand{\dif}{\partial}
\newcommand{\te}{T_{et}}
\newcommand{\den}{n}
\newcommand{\ti}{T_i}
\newcommand{\ad}{\Gamma}
\newcommand{\etae}{\eta_e}
\newcommand{\etai}{\eta_i}
\newcommand{\trap}{f_t}
\newcommand{\epsn}{\epsilon_n}
\newcommand{\dia}{{\omega_*}}
\newcommand{\wde}{\omega_{de}}
\newcommand{\wdi}{\omega_{di}}
\newcommand{\ky}{k_y}

\newcommand{\kperpsq}{k_\perp^2}

\newcommand{\gele}{g_e}
\newcommand{\gi}{g_i}
\newcommand{\ai}{\alpha_i}

\newcommand{\nk}{{n_k}}
\newcommand{\tek}{T_{ek}}
\newcommand{\nik}{{n_{ik}}}
\newcommand{\tik}{T_{ik}}
\newcommand{\fk}{{\phi_k}}
\newcommand{\tk}{{T_k}}
\newcommand{\psik}{{\psi_k}}
\newcommand{\freq}{\omega}
\newcommand{\wk}{{\omega_k}}
\newcommand{\gk}{{\gamma_k}}

\newcommand{\sek}{{s_{ek}}}
\newcommand{\sik}{{s_{ik}}}

\newcommand{\ratio}{\xi}

\newcommand{\coef}{\alpha_k}

\newcommand{\cpn}{\theta_k}
\newcommand{\cps}{{\zeta_k}}
\newcommand{\cpsi}{\chi_k}
\newcommand{\cpdif}{\psi_k}

\newcommand{\wres}{\omega_{\rm res}}

\newcommand{\ampone}{\alpha_k} 
\newcommand{\amptwo}{\beta_k}

\usepackage{amsmath}
\usepackage{graphicx}
\usepackage{hyperref}

\begin{document}
\maketitle

\begin{abstract}
Collisionless trapped-electron mode (CTEM) turbulence is an important contributor to heat and particle transport in fusion devices. The ITG/TEM fluid models are rarely treated analytically, due to the large number of transport channels involved, e.g. particle and ion/electron heat transport.
The $\nabla T_e$-driven CTEM fluid model [Anderson et al, \emph{Plasma Phys. Control. Fusion} 48, 651 (2006)] provides a simplified model, in the regime where the density gradient drive is negligeable compared to the electron temperature gradient drive ($\nabla T_e$). This provides an interesting model to study mechanisms associated to linear waves, such as crossphase dynamics, and its possible role in the formation of $E\times B$ staircase. Here, the  $\nabla T_e$-driven CTEM fluid model is rigourously derived from the more general ITG/TEM model, and its linear dynamics is first analyzed and compared with CTEM gyrokinetic simulations with bounce-averaged kinetic electrons, while nonlinear analysis is left for future work. Comparisons of linear ITG spectrum are also made with other analytical models.
\end{abstract}

\section{Introduction}
In magnetic fusion devices, electron heat transport and particle transport is partly due to collisionless trapped-electron mode (CTEM) turbulence, which is usually coupled to ion-temperature gradient (ITG) turbulence.
The CTEM is an instability due to the toroidal precession-drift resonance of trapped electrons in the low-collisionality regime \cite{Adam1976}. It is driven by electron temperature gradient and/or density gradient.
The Chalmers model \cite{Weiland2000, Nordman1990, Nilsson1990, Jarmen1987, Anderson2006} provides a simple yet predictive set of fluid models for ITG and CTEM turbulence. This set of models has been analyzed linearly. However, there remain certain aspects which are not fully understood, such as the crossphase dynamics responsible for the transport \cite{LeconteKobayashi2021, Sasaki2021, LeconteSingh2019, CYAn2017}. In this article, we first analyze the ITG-TEM model linearly - including the crossphase dynamics.
Focusing on the limit of $\nabla T_e$-driven CTEM, the two-field model of Ref. \cite{Anderson2006} is analyzed further
and compared to linear gyrokinetic simulations with bounce-averaged kinetic electrons \cite{KwonQiYi2017}.
The future goal is to analyze zonal electron temperature corrugations associated to zonal staircase with nonlinear simulations \cite{DifPradalier2015, DifPradalier2017, LeiQi2022}, its relation to zonal flows, and its possible relation to the transport crossphase \cite{LeconteKobayashi2021, Sasaki2021}. This is left for future work.

The rest of this article is organized as follows. In Section 2, we present the general ITG-TEM model used in this study, a linear analysis and general crossphase dynamics analysis is presented. The derivation of the ITG dispersion relation is reviewed and compared to linear gyrokinetic simulations. In section 3, we focus on the $\nabla T_e$-driven CTEM, a subset of the ITG-TEM model. The model is analyzed - including the detailed crossphase dynamics - and
compared with linear gyrokinetic simulations. In section 4, we discuss the results and present conclusions.

\section{Model}
We are interested in collisionless trapped electron modes (CTEM) for which the frequency verifies $k_\parallel v_{th,i} \ll \omega_k < \ky v_{th,e}$.
Although electron trapping is a toroidal phenomenon, we assume  a slab geometry, for simplicity.
We consider the following fluid model of collisionless trapped-electron mode turbulence, based on Nordman et al. \cite{Nordman1990}:
\begin{eqnarray}
\frac{\dif \den}{\dif t} & + & {\bf v}_E . \nabla \den + \trap v_{*e} \frac{\dif \phi}{\dif y} = - \epsn \gele v_{*e} \frac{\dif}{\dif y} ( \den - \trap \phi + \trap \te ),
\label{den00} \\
\frac{\dif \te}{\dif t} & + & {\bf v}_E . \nabla \te + \etae v_{*e} \frac{\dif \phi}{\dif y} = - \frac{\ad-1}{\trap} \epsn \gele v_{*e} \frac{\dif}{\dif y} \Big( \den - \trap \phi + (1+\frac{\ad}{\ad-1}) \trap \te \Big), \qquad
\label{temp00} \\
\Big[ \frac{\dif}{\dif t} & + & {\bf v}_E . \nabla \Big] \Big[ (1- \trap) \tilde \phi - \nabla_\perp^2 \phi \Big] + (1- \trap + \frac{1+\etai}{\tau} \nabla_\perp^2  ) v_{*e} \frac{\dif \phi}{\dif y} = \nonumber\\
& & \epsn \gi v_{*e} \frac{\dif}{\dif y} \Big[ (1- \trap)( \gele / \gi+ 1 / \tau)\phi + \ti + (\gele / \gi + 1/ \tau) \den + (\gele / \gi) \trap \te \Big],
\label{cb00} \\
\frac{\dif \ti}{\dif t} & + & {\bf v}_E . \nabla \ti  =
- \frac{v_{*e}}{\tau} \Big[ \etai + \frac{\ad -1}{\tau} (1+ \etai + \tau) \nabla_\perp^2  \Big] \frac{\dif \phi}{\dif y} \nonumber\\
 & + &  \frac{(\ad-1)\epsn \gi v_{*e}}{\tau} \frac{\dif}{\dif y} \Big[ \frac{1}{\tau} (1 - \trap + (\gele / \gi) \tau) \fk
+ (1+ \frac{\ad}{\ad-1}) \ti + \frac{\den}{\tau} \Big] \nonumber\\
 & + & \frac{(\ad-1) \epsn \gi v_{*e}}{\tau^2} \nabla_\perp^2 \frac{\dif}{\dif y} 
 \Big[  (1+\tau) \phi + \frac{\tau}{1- \trap} T_i + \frac{1+ \tau}{1- \trap} n + \frac{\trap \tau}{1- \trap} \te \Big]
\label{ti00}
\end{eqnarray}
Equation (\ref{den00}) represents the conservation of effective electron density, Eq. (\ref{temp00}) is the electron heat balance, Eq. (\ref{cb00}) is charge balance, and Eq. (\ref{ti00}) is the ion heat balance.
Here, $n = n_{et} + \trap \tilde \phi$ is an effective density, with $n_{et} = \int d^3v ~h_e / n_0$ the trapped electron density \cite{LangParkerChen2008, HahmTang1996}. Note that this convention for trapped electrons is a different convention than the one used in the Chalmers model. The quantity $\phi$ denotes the electric potential, $\tilde \phi = \phi - \langle \phi \rangle$, with $\langle \cdot \rangle = \frac{1}{L_y}\int dy $ the flux surface average, and ${\bf v}_E = \hat z \times \nabla \phi$ denotes the $E \times B$ drift. The electric potential is normalized as $\frac{e \phi}{T_{e0}} \to \phi$, with $T_{e0}$ a reference electron temperature. The quantity $\nabla_\perp^2 = \frac{\dif^2}{\dif x^2} + \frac{\dif^2}{\dif y^2}$ is the perpendicular Laplacian, and $x, y, z$ denote the local radial, poloidal and toroidal directions in a fusion device. Time is normalized as $(c_s/L_n) t \to t$, with $c_s = \sqrt{T_e} / m_i$ the sound speed, and space is normalized by the gyroradius at the electron temperature $\rho_s = c_s / \omega_{c,i}$, with $\omega_{c,i} = eB/m_i$ the ion gyrofrequency. The quantity $\tau = T_{e0} / T_{i0}$ is the temperature ratio, and $\ad = 5/3$ is the adiabatic index.
The parameter $\trap = \sqrt{\epsilon}$ is the trapped-electron fraction, with $\epsilon = a/R$ the inverse aspect ratio, and terms proportional to $\epsn = 2 L_n  / R$ denote magnetic drift effects modeling magnetic drift for ions and toroidal precession-drift for trapped-electrons. The  parameter $\gele(s) = \frac{1}{4} + \frac{2}{3} s$ represents the effect of magnetic shear on the precession frequency of trapped-electrons, and $\gi(s) = \frac{2}{3} + \frac{5}{9} s - \epsilon$ is its effect on the magnetic drift for ions \cite{AndersonNordman2006, Garbet2003}.
Since we are interested in frequencies $\omega \gg k_\parallel c_s$, where $c_s$ is the sound speed, we neglect parallel ion motion.

\subsection{Schr$\ddot {\bf o}$dinger-like equation}
In this section, we neglect magnetic shear effects, i.e. we assume $\gi=\gele=1$, for simplicity.
Based on Ref. \cite{ZhouZhuDodin2019}, the linearization of the system (\ref{den00} - \ref{ti00}) can be written in the form of the following 4-component Schr$\ddot {\rm o}$dinger-like equation:
\begin{equation}
i \frac{\dif}{\dif t}
\begin{bmatrix}
1 & 0 & 0 &0 \\
0 & 1 & 0 & 0 \\
0 & 0 & 1- \trap + \kperpsq & 0 \\
0 & 0 & 0 & 1
\end{bmatrix}
\begin{bmatrix}
\nk \\
\tek \\
\fk \\
\tik
\end{bmatrix}
= H
\begin{bmatrix}
\nk \\
\tek \\
\fk \\
\tik
\end{bmatrix},
\end{equation}
where $H$ denotes the Hamiltonian matrix, given by:
\begin{equation}
H =
\begin{bmatrix}
a & a & b & 0 \\
c & d & f & 0 \\
g & -a & h & -a \\
j & k & l & m
\end{bmatrix},
\end{equation}
where $a-m$ are real-valued coefficients given by:
\begin{eqnarray}
a= \wde, \quad b = \trap(\dia - \wde),
\label{expr1} \\
c = \frac{2}{3} \wde, \quad d= \frac{7}{3} \wde, \quad f= \trap (\etae \dia - \frac{2}{3} \wde),
\label{expr2} \\
g= - (1+\frac{1}{\tau}) \wde, \quad h= (1- \trap - \frac{1+ \etai}{\tau} \kperpsq) \dia - (1- \trap) (1+\frac{1}{\tau}) \wde,
\label{expr3} \\
j = - \frac{2}{3 \tau^2} (1-\frac{1+ \tau}{1- \trap} \kperpsq) \wde, \quad k= \frac{2}{3 \tau} \frac{\kperpsq}{1- \trap} \wde,  \\
l= \frac{1}{\tau} \Big[ (\etai - \frac{2}{3 \tau} (1+ \etai + \tau) \kperpsq ) \dia - \frac{2}{3 \tau} (1- \trap + \tau - (1+ \tau) \kperpsq) \wde  \Big],
\label{expr4} \\
m= - ( \frac{7}{3} - \frac{2}{3} \frac{\kperpsq}{1- \trap} ) \frac{\wde}{\tau}
\label{expr5}
\end{eqnarray}
To obtain the `normal coordinates', i.e. the eigenvectors associated to this system, one needs to diagonalize the $H$ matrix. 

Later on, we will focus on the `pure' CTEM turbulence, i.e. $\nabla T_e$-driven CTEM. That is, considering quasi-neutrality $n_i = n + (1-\trap) [\phi - \langle \phi \rangle]$, we assume that the mode frequency resonates with the precession-drift frequency, i.e. $\wk \sim \frac{5}{3} \wde$.  Hence, we use the approximation:
\begin{equation}
|n_i| \ll |n|, |\phi - \langle \phi \rangle|
\end{equation}
In this regime, the electron dynamics decouples from the ion dynamics, and the quasi-neutrality condition reduces to \cite{Anderson2006} :
\begin{equation}
n \simeq - (1- \trap) [\phi - \langle \phi \rangle]
\label{approx1}
\end{equation}
Hence, in this regime, the effective density fluctuations are proportional to the electric potential fluctuations, and there is no turbulent particle transport.
Using approximation (\ref{approx1}), the system (\ref{den00}-\ref{ti00}) reduces to the following pure CTEM model:
\begin{eqnarray}
\frac{\dif \tilde \phi}{\dif t} & + & {\bf v}_E . \nabla \tilde \phi - \ratio v_{*e} \frac{\dif \phi}{\dif y} = - \epsn v_{*e} \frac{\dif}{\dif y} [(1+ \ratio) \phi - \ratio \te ]
\label{den0} \qquad\\
\frac{\dif \te}{\dif t} & + & {\bf v}_E . \nabla \te + \etae v_{*e} \frac{\dif \phi}{\dif y} = \frac{2}{3 \ratio} \epsn v_{*e} \frac{\dif}{\dif y} [(1+ \ratio) \phi - \frac{7}{2} \ratio \te ],
\label{temp0}
\end{eqnarray}

with $\tilde \phi = \phi - \langle \phi \rangle$ and $\ratio= \trap / (1- \trap)$. 

For now, let us carry on the analysis of the full system (\ref{den00}-\ref{ti00}).


\subsection{Linear analysis}
Linearizing the system (\ref{den00}-\ref{ti00}), one obtains:
\begin{eqnarray}
\freq \nk & = & a (\nk+\tek) +b \fk,
\label{lin-den1} \\
\freq  \tek & = & c \nk + d \tek +f \fk,
\label{lin-te1} \\
(1- \trap + \kperpsq) \freq \fk & = & g \nk + h \fk -a (\tek+\tik),
\label{lin-cb1} \\
\freq \tik & = & j \nk + k \tek + l \fk + m \tik,
\label{lin-ti1}
\end{eqnarray}
where the coefficients $a-m$ are given by expressions $(\ref{expr1} - \ref{expr5})$.

After some algebra, one obtains the following ITG-TEM linear dispersion relation:
\begin{align}
\frac{ (\dia + \tau \wdi) (\freq - \frac{5}{3} \wdi ) - \kperpsq ( \freq + \frac{1+ \etai}{\tau} \dia) (\freq -\frac{5}{3} \wdi) + (\etai -\frac{7}{3} +\frac{5}{3} \epsn) \dia \wdi }{N_i} = \notag\\
\trap \frac{ (\dia - \wde) (\freq - \frac{5}{3} \wde) +(\etae - \frac{2}{3}) \dia \wde }{N_e} + 1 - \trap,
\label{lin-disp0}
\end{align}
where the denominators are $N_i = \freq^2 - \frac{10}{3} \omega_{di}\freq + \frac{5}{3} \omega_{di}^2$ and $N_e = \freq^2 - \frac{10}{3} \wde \freq + \frac{5}{3} \wde^2$.
Details of the derivation are given in Appendix. Eq. (\ref{lin-disp0}) is a quartic dispersion relation which describes two coupled modes, an electron mode (CTEM rotating in the electron diamagnetic direction), and an ion mode (ITG rotating in the ion diamagnetic direction) \cite{Anderson2006}. As we focus here on trapped electron mode turbulence, we assume $\eta_i \ll \etae$, and thus the electron branch (CTEM branch) is dominant. It resonates close to the toroidal precession drift frequency $\wk \propto \wde$. In the $\nabla T_e$-driven CTEM regime, $|N_i| \gg |N_e|$. Hence the dispersion relation reduces to a quadratic, and the CTEM mode resonates at the frequency $\wk = \wres = \frac{5}{3} \wde$. We are talking here about a fluid resonance, which is an approximation of the kinetic resonance \cite{Weiland2000}.

From the linearized Eqs. (\ref{lin-den1}, \ref{lin-te1}, \ref{lin-cb1}, \ref{lin-ti1}), one obtains - after some algebra - the following system:
\begin{eqnarray}
[ \freq - \wde + \Lambda_k^{\rm TEM} ] \nk - \trap \wde (s_{ek} + \frac{2}{3 \trap} \nk)  - \Lambda_k^{\rm TEM} \nik & = & 0,
\label{lin-den3} \qquad\\
\freq s_{ek} - \frac{5}{3} \wde (s_{ek} + \frac{2}{3 \trap} \nk) + (\etae - \frac{2}{3}) \frac{\dia}{1- \trap} \nk - (\etae - \frac{2}{3}) \frac{\dia}{1- \trap} \nik & = & 0,
\label{lin-se3} \\
\freq s_{ik}  - \frac{5}{3} \wdi (s_{ik} + \frac{2}{3 \tau} \nik) + (\etai - \frac{2}{3}) \frac{\dia / \tau}{1 - \trap} \nk - (\etai - \frac{2}{3}) \frac{\dia / \tau}{1 -\trap} \nik & = & 0,
\label{lin-si3} \\
\Lambda_k^{\rm ITG} \nk
+ \Big[ \freq - \wdi - \Lambda_k^{\rm ITG} \Big] \nik - \tau \wdi (s_{ik} + \frac{2}{3 \tau} \nik) & = & 0,
\label{lin-ni3}
\end{eqnarray}
where we used quasi-neutrality to express the electric potential: $\fk = \frac{1}{1- \trap} (\nik - \nk)$, the parameters $\Lambda_k^{\rm TEM} = \ratio (\dia - \wde)$ and $\Lambda_k^{\rm ITG} (\freq) = \frac{1}{1 - \trap} \Big[ (1- \frac{1+ \etai}{\tau} \kperpsq) \dia + \tau \wdi - \kperpsq \freq \Big]$ were defined, and $\xi = \trap / (1- \trap)$. The following quantities were introduced:
\begin{equation}
\sik = \tik - \frac{2}{3 \tau} \nik, \quad {\rm and} \quad \sek = \tek - \frac{2}{3 \trap} \nk,
\end{equation}
where $s_{ik}$ ($s_{ek}$) is the thermodynamic entropy for ions (electrons), as described for ITG in Refs. \cite{DiamondLect2021, GarbetDubuitAsp2005}.

In matrix form, the ITG-TEM system takes the simple form:
\begin{equation}
\begin{bmatrix}
\freq - \frac{5}{3} \wde + \Lambda_k^{\rm TEM} & - \trap \wde & - \Lambda_k^{\rm TEM} & 0 \\
(\etae - \frac{2}{3}) \frac{\dia}{1- \trap} - \frac{10}{9 \trap} \wde & \freq - \frac{5}{3} \wde & - (\etae - \frac{2}{3}) \frac{\dia}{1- \trap} & 0 \\
\Lambda_k^{\rm ITG}(\freq) & 0 & \freq - \frac{5}{3} \wdi - \Lambda_k^{\rm ITG}(\freq) & - \tau \wdi \\
(\etai - \frac{2}{3}) \frac{\dia / \tau }{1- \trap} & 0 & - \Big[  (\etai - \frac{2}{3}) \frac{\dia / \tau}{1 -\trap} + \frac{10}{9 \tau} \wdi \Big] &
\freq - \frac{5}{3} \wdi
\end{bmatrix}
\begin{bmatrix}
\nk \\
\sek \\
\nik \\
\sik
\end{bmatrix}
= 0
\label{matrix1}
\end{equation}

This form of the ITG-TEM system clearly shows that the system is almost block-diagonal, with the coupling between the two branches occuring only through the effective electron density $\nk$ and the ion density $\nik$. The two branches decouple in the following two limits: i) The toroidal ITG mode is recovered in the limit of negligeable trapped electron fraction $\trap \to 0$ corresponding to Boltzmann electrons, for which the effective electron density is negligeable $|\nk| \ll |\nik|, |\sik|$. It corresponds to the lower block-diagonal. Note that the total electron density in this case is simply $n_{ek} = \nik \simeq \fk$. ii) The pure-CTEM mode, i.e. $\nabla T_e$-driven CTEM is recovered in the limit $|\nik| \ll |\nk|, |\sek|$. It corresponds to the upper block-diagonal.

An alternative way to represent the system is in terms of the ion gyrocenter density $n_i^{\rm GC} = n_i - \nabla_\perp^2 \phi$ instead of ion density $n_i$. Then, the electric potential is given by $\fk = \frac{1}{1- \trap + \kperpsq} (n_{ik}^{\rm GC}- \nk)$, and the system can be written in the form of the following Schr$\rm \ddot o$dinger-like equation:
\begin{equation}
i \frac{\dif}{\dif t} \Psi
=
\begin{bmatrix}
\frac{5}{3} \wde - \Lambda_k^{\rm TEM} & \trap \wde & \Lambda_k^{\rm TEM} & 0 \\
- \Big[ (\etae - \frac{2}{3}) \frac{\dia}{1- \trap + \kperpsq} - \frac{10}{9 \trap} \wde \Big] & \frac{5}{3} \wde & (\etae - \frac{2}{3}) \frac{\dia}{1- \trap + \kperpsq} & 0 \\
- \Lambda_k^{\rm GC} & 0 & \frac{5}{3} \wdi + \Lambda_k^{\rm GC} & \tau \wdi \\
- (\etai - \frac{2}{3}) \frac{\dia / \tau}{1- \trap + \kperpsq} & 0 & (\etai - \frac{2}{3}) \frac{\dia / \tau}{1 -\trap + \kperpsq} + \frac{10}{9 \tau} \wdi &
\frac{5}{3} \wdi
\end{bmatrix}
\Psi,
\label{matrix1}
\end{equation}
where $\Psi = [\nk, \sek,  n_{ik}^{\rm GC}, \sik]^T$ is the eigenvector, the coefficient $\Lambda_k^{\rm ITG} (\freq)$ is now replaced by $\Lambda_k^{\rm GC} =  \frac{1}{1 - \trap + \kperpsq} \Big[ (1- \frac{1+ \etai}{\tau} \kperpsq) \dia + \tau \wdi - \frac{5}{3} \kperpsq \wdi \Big]$ which does not depend on the complex frequency $\freq$, and $\Lambda_k^{\rm TEM}$ is now defined as $\Lambda_k^{\rm TEM} = \ratio_k (\dia - \wde)$, where the trapping parameter $\ratio_k = \trap / (1- \trap + \kperpsq)$ now depends on the squared wavenumber $\kperpsq$.


\subsection{Crossphase dynamics}
To analyze the crossphase dynamics, it is convenient to re-write Eqs. (\ref{lin-den3}), (\ref{lin-se3}) and (\ref{lin-si3}) in terms of $\nk$, $\sek$, $\sik$ and the electric potential $\fk$:

\begin{eqnarray}
\Big[ \freq - \frac{5}{3} \wde \Big] \nk -  \trap \wde \sek - \trap (\dia- \wde) \fk  & = & 0
 \label{lin-den4} \qquad\\
(\freq - \frac{5}{3} \wde) \sek - (\etae - \frac{2}{3}) \dia \fk - \frac{10}{9 \trap} \wde \nk  & = & 0
 \label{lin-se4} \\
(\freq - \frac{5}{3} \wdi ) \sik - \Big[ (\etai - \frac{2}{3 \tau}) \dia + (1- \trap) \frac{10}{9 \tau} \wdi  \Big]  \fk  - \frac{10}{9 \tau} \wdi \nk  & = & 0
 \label{lin-si4}
\end{eqnarray}

Let us define the 3 crossphases:
\begin{equation}
\cpn = {\rm arg} ( \frac{\fk}{\nk} ), \quad \cps = {\rm arg} ( \frac{\fk}{\sek} ), \quad \cpsi = {\rm arg} ( \frac{\fk}{\sik} )
\end{equation}

With these definitions, the turbulent particle flux is $\Gamma = \sum_k \ky |\nk| |\fk| \sin \cpn$, and the electron/ion entropy fluxes are $\sum_k \ky |\sek| |\fk| \sin \cps$ and $\sum_k \ky |\sik| |\fk| \sin \cpsi$, respectively. The ion and electron entropy fluxes are related to the ion and electron heat fluxes $Q_{e,i}$ via $\sum_k \ky {\rm Im}~ \phi_k^* \sek = Q_e - \frac{2}{3 \trap} \Gamma$, and $\sum_k \ky {\rm Im}~ \phi_k^* \sik = Q_i - \frac{2}{3 \tau} \Gamma$, where use has been made of the quasi-neutrality condition $\nik = \nk +(1-\trap) \fk$.

We use the following ansatz:
\begin{equation}
\begin{bmatrix}
\fk  \\
\nk  \\
\sek \\
\sik 
\end{bmatrix}
=
\begin{bmatrix}
|\fk|  \\
|\nk| e^{-i \cpn}  \\
|\sek| e^{-i \cps} \\
|\sik| e^{-i \cpsi} 
\end{bmatrix}
e^{-i \wk t},
\label{ansatz1}
\end{equation}
where $\wk$ is the linear mode frequency.

Replacing $\nik, \nk , \sek$ and $\sik$ in Eqs. (\ref{lin-den4}-\ref{lin-si4}), using the ansatz (\ref{ansatz1}), one obtains, after some algebra, the following system:
\begin{align}
\gk e^{-i \cpn}|\nk| - i |\nk| e^{-i \cpn} \Big[ \frac{\dif \cpn}{\dif t} + \wk \Big]  = - \frac{5}{3} i \wde |\nk| e^{-i \cpn} \nonumber\\
- i \trap \wde |\sek| e^{-i \cps} 
- i \trap (\dia - \wde) |\fk|,
\label{lin-den5} \\
\gk e^{-i \cps} |\sek| - i |\sek| e^{-i \cps} \Big[ \frac{\dif \cps}{\dif t} + \wk \Big] = - \frac{5}{3} i \wde |\sek| e^{-i \cps} \nonumber\\
- i (\etae - \frac{2}{3}) \dia |\fk| - i \frac{10}{9 \trap} \wde |\nk| e^{-i \cpn},
\label{lin-se5}  \\
\gk e^{-i \cpsi} |\sik| - i |\sik| e^{-i \cpsi} \Big[ \frac{\dif \cpsi}{\dif t} + \wk \Big] = - \frac{5}{3} i \wdi |\sik| e^{-i \cpsi} \nonumber\\
- i \Big[ (\etai - \frac{2}{3 \tau} ) \dia + (1- \trap) \frac{10}{9 \tau} \wdi \Big] |\fk|
- \frac{10}{9 \tau} \wdi |\nk| e^{-i \cpn},
\label{lin-si5}
\end{align}
where we used $\dif_t |u_k| = \gk |u_k|$, for $u_k = \nk, \sek, \sik$. It is apparent that Eqs. (\ref{lin-den5},\ref{lin-se5}) can be solved independently from Eq. (\ref{lin-si5}). Physically, the crossphase $\cpsi$ responsible for ion thermal transport does not influence the crossphases $\cpn$ and $\cps$ associated to electron transport. We will thus first solve for the electron crossphases. One can separate the real and imaginary parts of Eqs. (\ref{lin-den5},\ref{lin-se5}). The real part yields two relations between the crossphases $\cpn$, $\cps$ and the linear growth-rate:
\begin{eqnarray}
\gk & = & \trap (\dia - \wde) \ampone \sin \cpn + \trap \wde \frac{\ampone}{\amptwo} \sin \cpdif ,
\label{amp-n0} \\
\gk & = & - \frac{10}{9 \trap} \wde \frac{\amptwo}{\ampone} \sin \cpdif + (\etae - \frac{2}{3}) \dia \amptwo \sin \cps ,
\label{amp-s0}
\end{eqnarray}
where $\gk = |\nk|^{-1} \dif_t |\nk|= |\sek|^{-1} \dif_t |\sek| = |\sik|^{-1} \dif_t |\sik|$ is the linear growth-rate, and $\cpdif = \cpn - \cps$ is the crossphase mismatch responsible for transport decoupling between particle transport v.s. electron heat transport. The quantities $\ampone$ and $\amptwo$ are the two amplitude ratios defined as:
\begin{equation}
\ampone = \frac{|\phi_k|}{|\nk|}, \quad \amptwo = \frac{|\phi_k|}{|\sek|},
\end{equation}
and $\ampone / \amptwo = |\sek| / |\nk| $ is the third amplitude ratio.

\begin{figure}
\includegraphics[width=0.5\linewidth]{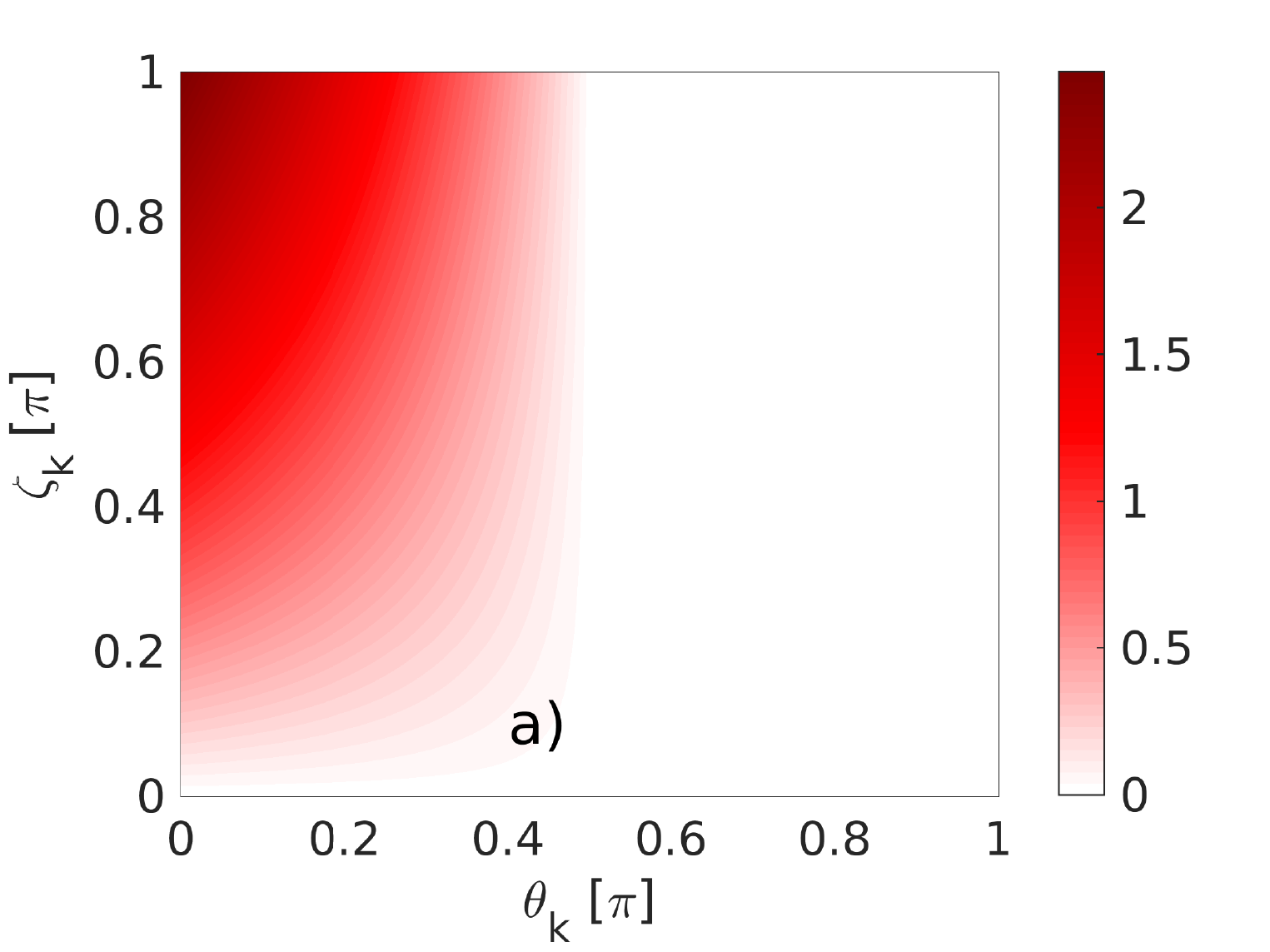}\includegraphics[width=0.5\linewidth]{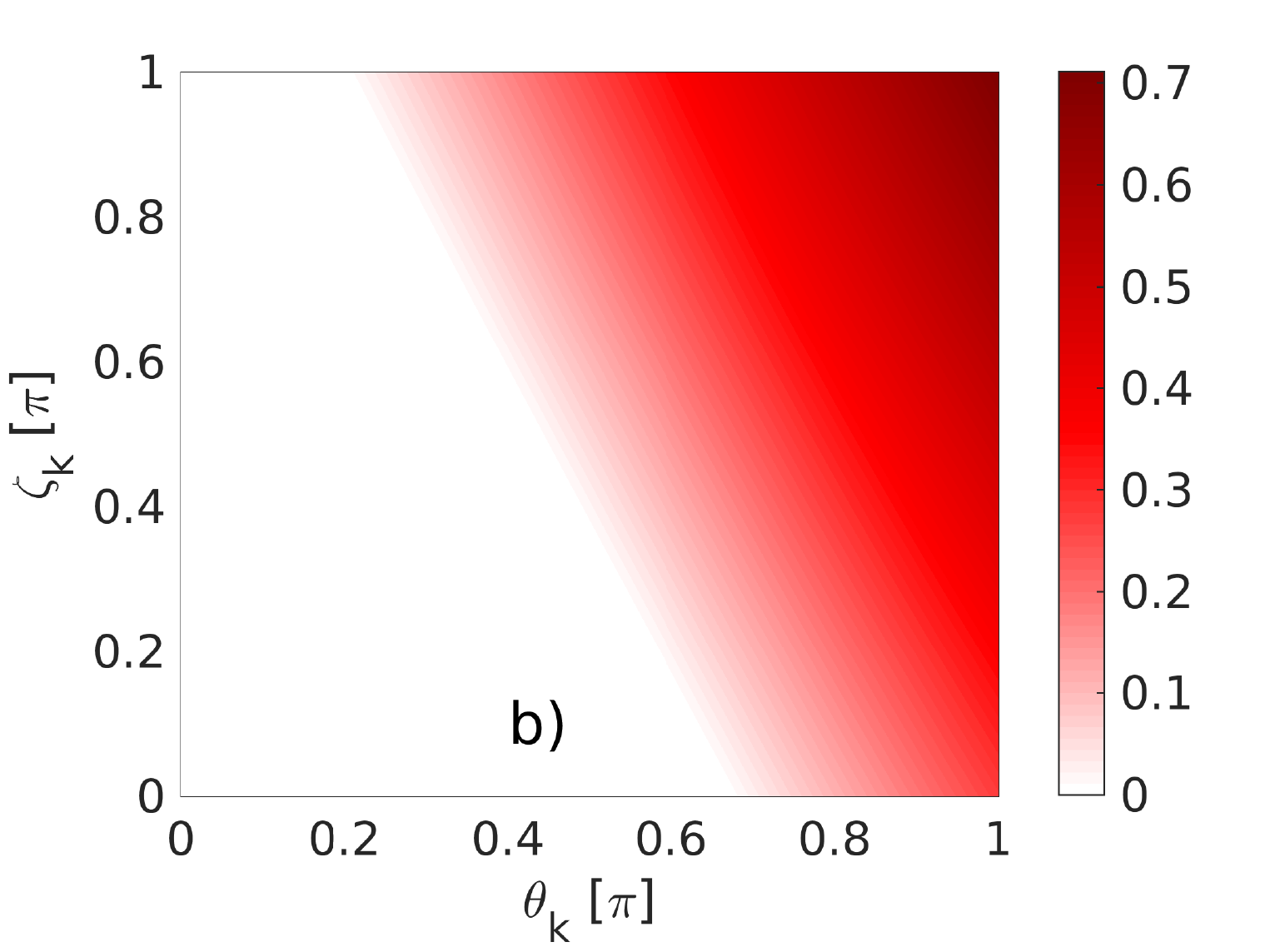}
\caption{ Amplitude ratios for ITG/TEM: a) inverse amplitude ratio $A_k = \frac{1}{\ampone}$ and b) $B_k = \frac{1}{\amptwo}$, expressions (\ref{invampratio1},\ref{invampratio2}) v.s. $\cpn$ and $\cps$, for the parameters $\etae=2$, $\epsn=0.8$ and $\gk/ \dia=1$.}
\label{fig-ampratio}
\end{figure}

The imaginary part of Eqs. (\ref{lin-den5},\ref{lin-se5}) yields the following crossphase dynamics.
\begin{eqnarray}
\frac{\dif \cpn}{\dif t} & = & \wres - \wk + \trap (\dia - \wde) \ampone \cos \cpn + \trap \wde \frac{\ampone}{\amptwo} \cos \cpdif,
\label{cpn0} \\
\frac{\dif \cps}{\dif t} & = & \wres - \wk + \frac{10}{9 \trap} \wde \frac{\amptwo}{\ampone} \cos \cpdif + (\etae - \frac{2}{3}) \dia \amptwo \cos \cps, 
\label{cps0} \\
\frac{\dif \cpdif}{\dif t} & = & \Big(  \trap \frac{\ampone}{\amptwo} - \frac{10}{9 \trap} \frac{\amptwo}{\ampone} \Big) \wde \cos \cpdif + \trap (\dia - \wde) \ampone \cos \cpn - (\etae - \frac{2}{3}) \dia \amptwo \cos \cps,
 \label{cp-dif0} \qquad
\end{eqnarray}
where $\wres = \frac{5}{3} \wde$ is the resonance frequency. Here, Eq. (\ref{cp-dif0}) is the difference of Eq. (\ref{cpn0}) and Eq. (\ref{cps0}).
Considering the amplitude ratio system (\ref{amp-n0},\ref{amp-s0}), it is convenient to define the inverse amplitude ratios $A_k = 1/ \ampone$ and $B_k = 1 / \amptwo$. After some algebra, one obtains the following matrix system:
\begin{equation}
\begin{bmatrix}
\gk & \trap \wde \sin \cpdif  \\
- \frac{10}{9 \trap} \wde \sin \cpdif  & \gk
\end{bmatrix}
\begin{bmatrix}
A_k \\
B_k
\end{bmatrix}
=
\begin{bmatrix}
\trap (\dia - \wde) \sin \cpn  \\
(\etae - \frac{2}{3}) \dia \sin \cps 
\end{bmatrix}
\end{equation}
Inverting the matrix yields after some algebra:
\begin{eqnarray}
\frac{1}{\ampone} & = & \frac{ \trap [ (1 - \epsn) \dia / \gk] \sin \cpn - \trap (\etae - \frac{2}{3}) (\epsn \dia^2 / \gk^2) \sin \cps \sin \cpdif }{ 1 + \frac{10}{9} \epsn^2 (\dia^2 / \gk^2)  \sin^2 \cpdif },
\label{invampratio1} \\
\frac{1}{\amptwo} & = & \frac{ (\etae - \frac{2}{3}) (\dia / \gk) \sin \cps + \frac{10}{9} [ (1 - \epsn) \epsn \dia^2 / \gk^2] \sin \cpn \sin \cpdif }{ 1 + \frac{10}{9} \epsn^2 (\dia^2 / \gk^2)  \sin^2 \cpdif }, \quad
\label{invampratio2}
\end{eqnarray}
The inverse amplitude ratios $A_k$ and $B_k$, expressions (\ref{invampratio1},\ref{invampratio2}) are plotted v.s. $\cpn$ and $\cps$, for the parameters $\etae=2$, $\epsn=0.8$ and $\gk/ \dia=1$ [Fig.\ref{fig-ampratio}]. Values on the forbidden domains - since $A_k$ and $B_k$ are amplitudes, they must be positive - are set to zero for clarity.

Alternatively, this can also be written:
\begin{eqnarray}
\ampone & = & \frac{\gk / \dia}{ \trap (1 - \epsn) \sin \cpn + \frac{1}{\amptwo} \trap \epsn \sin \cpdif }
\label{amp-n1} \\
\amptwo & = & \frac{ \gk / \dia }{ (\etae - \frac{2}{3} ) \sin \cps - \frac{1}{\ampone} \cdot \frac{10}{9 \trap} \epsn \sin \cpdif }
\label{amp-s1}
\end{eqnarray}
It is clear from Eqs. (\ref{amp-n1},\ref{amp-s1}) that the two amplitude ratios $\ampone$ and $\amptwo$ are coupled.
One identifies two opposite regimes: $\frac{1}{\amptwo} \ll 1$, corresponding to negligeable electron heat transport, and $\frac{1}{\ampone} \ll 1$ corresponding to negligeable particle transport. In the following, we will focus on the latter case, since we assume $\epsn \sim 1$.
In the regime $\frac{1}{\ampone} \ll 1$, the turbulence corresponds to $\nabla T_e$-driven CTEM. In this regime, one may neglect the coupling of the amplitude ratio $\amptwo$ to $\ampone$, i.e. one uses the approximation:
\begin{equation}
\amptwo \simeq \frac{\gk}{ (\etae - \frac{2}{3}) \dia \sin \cps}
\label{amp-n2}
\end{equation}
Then, the dynamics of the crossphase $\cps$ decouples from the other crossphases $\cpn$ and $\cpdif$, and reduces to:
\begin{equation}
\frac{\dif \cps}{\dif t} \simeq - ( \wk - \wres ) + \gk \cot \cps,
\label{cps1}
\end{equation}
where $\cot \cps = 1 / \tan \cps$ is the cotangent of the crossphase.


\subsection{ITG limit}
Linearizing the original ITG-TEM system (\ref{den00}-\ref{ti00}), one obtains:
\begin{eqnarray}
- i \freq \nk + i \trap \dia \fk = - i \epsn \gele \dia( \nk - \trap \fk + \trap \tek),
\label{lin-den2} \\
-i \freq \tek + i \etae \dia \fk = - \frac{2}{3 \trap} i \epsn \gele \dia (\nk - \trap \fk + \frac{7}{2} \trap \tek),
\label{lin-temp2} \qquad\\
-(1- \trap + \kperpsq) i \freq \fk + i (1- \trap - \frac{1+ \etai}{\tau} \kperpsq) \dia \fk = \nonumber\\
i \epsn \gele \dia [ (1- \trap)(1 + \gi/ (\gele\tau))\fk + (\gi/\gele) \tik +(1+ \gi/ (\gele\tau))\nk + \trap \tek ],
\label{lin-cb2} \\
-i \freq \tik = i \frac{\epsn \gele \dia}{\tau} \Big[ \frac{2}{3 \tau} (1 - \trap + \tau) \fk + \frac{2}{3 \tau} \nk + \frac{7}{3} \tik \Big]
- i \Big[ \etai - \frac{2}{3 \tau} (1+ \etai +\tau) \kperpsq \Big] \frac{\dia}{\tau} \fk \nonumber\\
- \frac{2}{3 \tau^2} i \kperpsq \epsn \gele \dia \Big[ (1+ \tau) \fk + \frac{\tau}{1- \trap} \tik + \frac{1+ \tau}{1- \trap} \nk + \frac{\tau \trap}{1- \trap} \tek \Big],
\label{lin-ti2} 
\end{eqnarray}
with $\dia = \ky v_{*e}$ the electron diamagnetic frequency.
The linear dispersion relation is best obtained by first transforming the system. Adding Eqs. (\ref{lin-den2}) and (\ref{lin-cb2}) yields the ion continuity equation:
\begin{equation}
- i \freq (n_{ik} + \kperpsq \fk ) + \Big( 1 -\frac{1+ \etai}{\tau} \kperpsq \Big) i \dia \fk = i \frac{\epsn \gi \dia}{\tau} \Big[ \Big( 1- \trap + \tau \Big) \fk + \nk + \tau \tik \Big],
\label{lin-ion2}
\end{equation}
where $n_{ik} = n_k + (1- \trap) \fk$ is the ion density perturbation set by quasi-neutrality (this is the particle - not gyrocenter - density).
In the ITG limit, i.e. without trapped-electrons $\trap \to 0$, the ion continuity equation and ion heat balance become:
\begin{eqnarray}
- i \freq (n_{ik} + \kperpsq \fk ) + \Big( 1 -\frac{1+ \etai}{\tau} \kperpsq \Big) i \dia \fk = i  \frac{\epsn \gi \dia}{\tau} \Big[ (1+\tau)\fk + \nik + \tau \tik \Big], \\
(-i \freq -i \frac{5}{3} \frac{\epsn \gi \dia}{\tau}) \tik + \frac{2}{3 \tau} i \freq \nik = \frac{2}{3 \tau} (\etai - \frac{2}{3}) i \dia \fk
\end{eqnarray}

\begin{figure}
\includegraphics[width=0.5\linewidth]{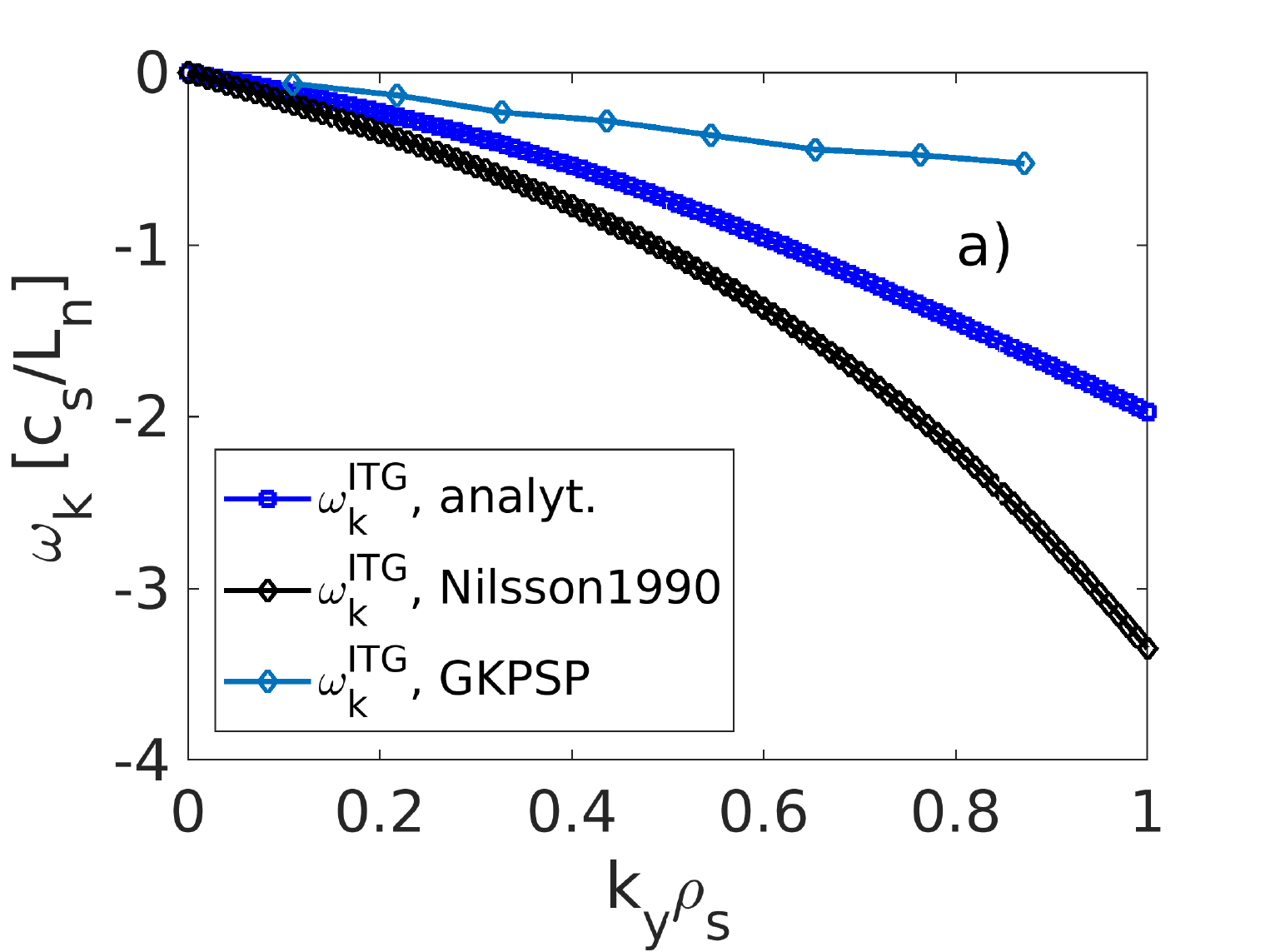}\includegraphics[width=0.5\linewidth]{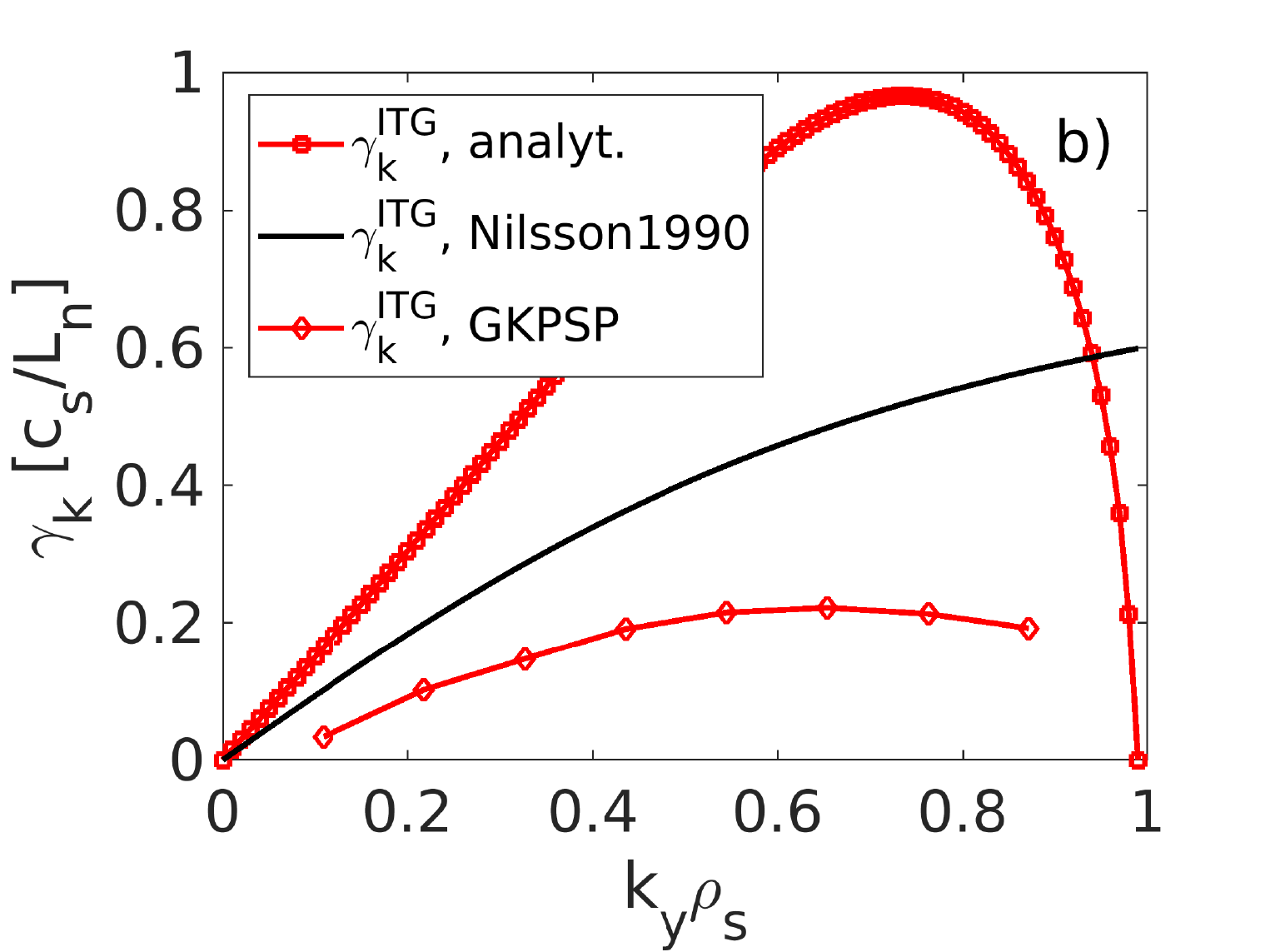}
\caption{Comparison of the analytical model with Nilsson \emph{et al.}1990 \cite{Nilsson1990} and with gyrokinetic simulations: a) ITG frequency and b) ITG growth-rate v.s. poloidal wavenumber $\ky$. The parameters are $\eta_i = 3$, $\epsn=1$ ($\frac{R}{L_n} =2$), $T_i/T_e=1$ and $\etae \simeq 0 $, with Boltzmann electrons. }
\label{fig-itg}
\end{figure}

\begin{figure}
\begin{center}
\includegraphics[width=0.5\linewidth]{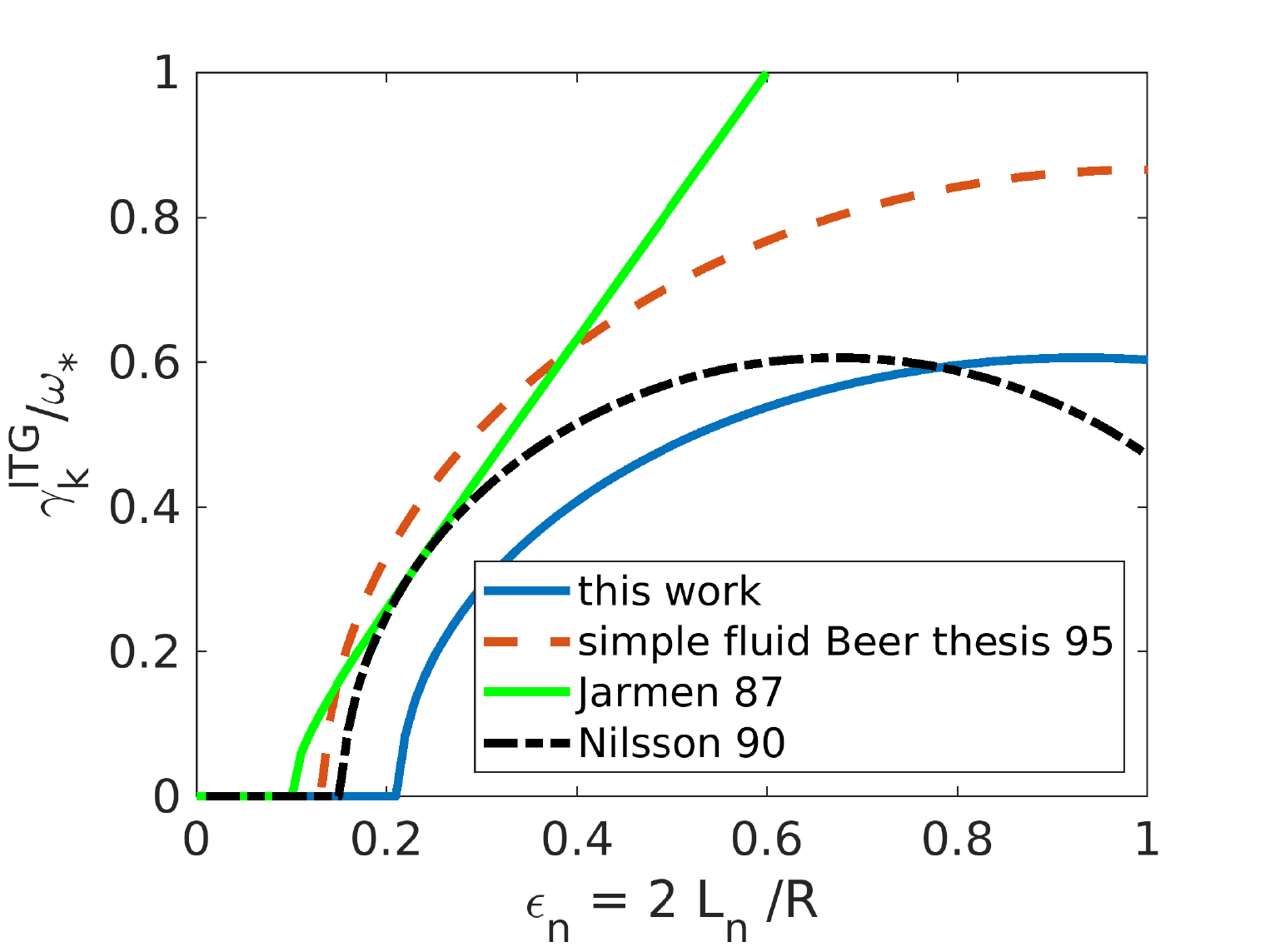}
\caption{ITG growth-rate v.s. $\epsn$ for the analytical model Eq. (\ref{growth-itg}) (blue) and comparison with three previous analytical models: the `simple fluid' model from Beer thesis \cite{Beer1995} (red), the result from Jarmen \emph{et al.} 1987 \cite{Jarmen1987} (green), and by Nilsson \emph{et al.} 1990 \cite{Nilsson1990} (black). Other parameters are $\eta_i=2$, $\kperpsq=0$ and $\tau_i=1$.}
\label{fig-itg-epsn}
\end{center}
\end{figure}

Defining $\tau_i = T_i / T_e = 1 / \tau$, the ion continuity Eq. (\ref{lin-ion2}) and ion heat balance Eq. (\ref{lin-ti1}) become, respectively:
\begin{eqnarray}
-i \freq \nik +(-i \freq - i \ai \ky) \kperpsq \fk + i \ky \fk - i \tau_i \epsn \gi \ky [ (1+\frac{1}{\tau_i})\fk + \nik+ \frac{\tik}{\tau_i} )] = 0,
\label{lin-itg11} \\
-i \freq \tik - i \ky \frac{5}{3} \tau_i \epsn \gi \tik + (\etai - \frac{2}{3}) i \tau_i \ky \fk + \frac{2}{3} i \tau_i \freq \nik = 0,
\label{lin-itg12}
\end{eqnarray}
with $\ai = \tau_i (1+ \etai)$. Here, the fields have been further normalized as $\nik\to \frac{L_n}{\rho_s} \nik$, and same for $\tik$.
For ITG, electrons are Boltzmann $\nik = \fk$. From Eq. (\ref{lin-itg12}), one obtains after some algebra the following linear response of ion temperature to potential :
\begin{equation}
\tik = \frac{(\etai-\frac{2}{3}) \tau_i \ky + \frac{2}{3} \tau_i \freq}{\freq + \frac{5}{3} \tau_i \epsn \gi \ky} \fk
\label{lin-itg22}
\end{equation}
Replacing $\tik$ in Eq. (\ref{lin-itg11}) using the response (\ref{lin-itg22}), one obtains - after some algebra - the following quadratic dispersion relation:
\begin{align}
A \freq^2 + B \freq + C=0,
\label{disprel-itg}
\end{align}
with real-valued coefficients given by $A=1+ \kperpsq$, $B = 2 \ky \Big[ \frac{5}{3} \tau_i \epsn \gi - \frac{1}{2} (1- \epsn \gi) + \frac{\kperpsq}{2} (\ai + \frac{5}{3} \tau_i \epsn \gi) \Big]$ and $C = \tau_i \epsn \gi \ky^2 \Big[ \etai - \frac{7}{3} +\frac{5}{3} (1+ \tau_i) \epsn \gi + \frac{5}{3} \ai \kperpsq \Big]$.
The solution to the ITG dispersion relation (\ref{disprel-itg}) is then:
\begin{align}
\freq_{1,2} \simeq - \frac{B}{2A} \pm \frac{1}{2A} \sqrt{B^2 - 4 AC}
\end{align}
The unstable branch has a growth-rate $\gk \simeq \frac{1}{ \sqrt{A} } \sqrt{C -\frac{B^2}{4A}} = \frac{1}{ \sqrt{A} } \sqrt{C -A(\frac{B}{2A})^2}$. 
After some algebra, the ITG frequency and growth-rate are given by:
\begin{eqnarray}
\wk^{\rm ITG} & \simeq & \frac{- \ky}{1+\kperpsq} \Big[ \frac{5}{3} \tau_i \epsn \gi - \frac{1}{2} (1- \epsn \gi) + \frac{\kperpsq}{2} (\ai + \frac{5}{3} \tau_i \epsn \gi) \Big],
\label{freq-itg} \\
\gk^{\rm ITG} & \simeq & \frac{\ky \sqrt{\tau_i \epsn \gi}}{ \sqrt{1+ \kperpsq} } \sqrt{\etai - \etai^c},
\label{growth-itg}
\end{eqnarray}
where
$\eta_i^c = \frac{1}{\tau_i \epsn \gi} (1+ \kperpsq) \frac{\wk^2}{\ky^2} + \frac{7}{3} - \frac{5}{3} (1+ \tau_i) \epsn \gi - \frac{5}{3} \ai \kperpsq$
is the linear ITG threshold, with $\wk = \wk^{\rm ITG}$ \cite{Anderson2002}. Note that here the growth-rate has a factor $\sqrt{1+ \kperpsq}$ in the denominator, compared to the factor $(1+ \kperpsq)$ in Ref. \cite{Anderson2002}. We believe there was a typo in \cite{Anderson2002}.  \\
The analytic expressions for the real ITG frequency (\ref{freq-itg}) and growth-rate (\ref{growth-itg}) are compared to Eqs. (13b,13c) in Nilsson \emph{et al.} \cite{Nilsson1990}, and to linear gyrokinetic simulations with the global gyrokinetic code GKPSP which includes fully gyrokinetic ions and bounce-averaged kinetic electrons \cite{KwonQiYi2017}. Here, Boltzmann electrons are used for the comparison. To take into account magnetic shear effects, the perpendicular wavenumber $k_\perp^2$ is replaced by its average over a ballooning trial function, yielding $\langle k_\perp^2 \rangle = [1+(\pi^2-7.5) s^2/3]\ky^2$ \cite{Nordman1993}. The parameters are $R/L_n = 2$, $R/L_{Ti}=6$, $R/L_{Te} \simeq 0$, and $T_i/T_e=1$. The safety factor is $q=1.34$ and the magnetic shear $s=0.75$. The ITG frequency given by expression (\ref{freq-itg})  is plotted v.s. $\ky$  (squares) and compared to the GKPSP result (diamonds) and to the analytical expression (13b) of Nilsson \emph{et al.} \cite{Nilsson1990} [Fig.\ref{fig-itg}a].
The frequency of the  ITG fluid model only shows qualitative similarity with the GKPSP data, but our analytical result shows better agreement with GKPSP data than Ref. \cite{Nilsson1990}, especially at large $\ky\rho_s$. This is probably due to the different treatment of ion FLR effects. \newline
The ITG growth-rate given by expression (\ref{growth-itg}) is plotted v.s. $\ky$ (squares) and compared to the Nilsson expression (13c) in Ref. \cite{Nilsson1990} (black), and to the GKPSP result (diamonds) [Fig.\ref{fig-itg}b].
The growth-rate also shows qualitative similarity, although the growth-rate of the fluid model is much larger than the GKPSP result and peaks at a slightly higher wavenumber. The result from Ref. \cite{Nilsson1990} seems to show closer agreement with the GKPSP data at small $\ky\rho_s$. However, the growth-rate peaks at $\ky\rho_s >1$, which is unphysical. This shows that the formula of Ref. \cite{Nilsson1990} is only valid for $\ky\rho_s \ll 1$. \newline
We also plot the ITG growth-rate - normalized to $\dia$ - v.s. the parameter $\epsn$ [Fig.\ref{fig-itg-epsn}], in the case $\eta_i=2$, $\kperpsq=0$ and $\tau_i=1$. Previous analytical results are also plotted, for comparison. In particular, in the Introduction section of Beer's PhD thesis \cite{Beer1995}, the ITG growth-rate from a `simple fluid' model is compared to results from the `3+1 Gyro-Landau-Fluid' model. We cannot compare our results to the latter, as we don't have access to this data, but a comparison is made with the `simple fluid' case. Note that here, $\epsn = 2 L_n /R$ as opposed to $L_n/R$ in Ref. \cite{Beer1995}. At the value $\epsn \sim 1$ - relevant for core plasmas - one can see that expression (\ref{growth-itg}) gives a lower growth-rate than the simple fluid model in Ref. \cite{Beer1995}, and the Jarmen et al. \cite{Jarmen1987} formula, which are more consistent with the `gyrofluid' and `kinetic' value of $\gamma/ \dia \simeq 0.4$ at $\epsn=1$ in Ref. \cite{Beer1995} (not shown). The Nilsson formula from Ref. \cite{Nilsson1990} has even better agreement with the latter results, compared to expression (\ref{growth-itg}).

\section{$\nabla T_e$-driven CTEM limit}

\begin{figure}
\begin{center}
\includegraphics[width=0.5\linewidth]{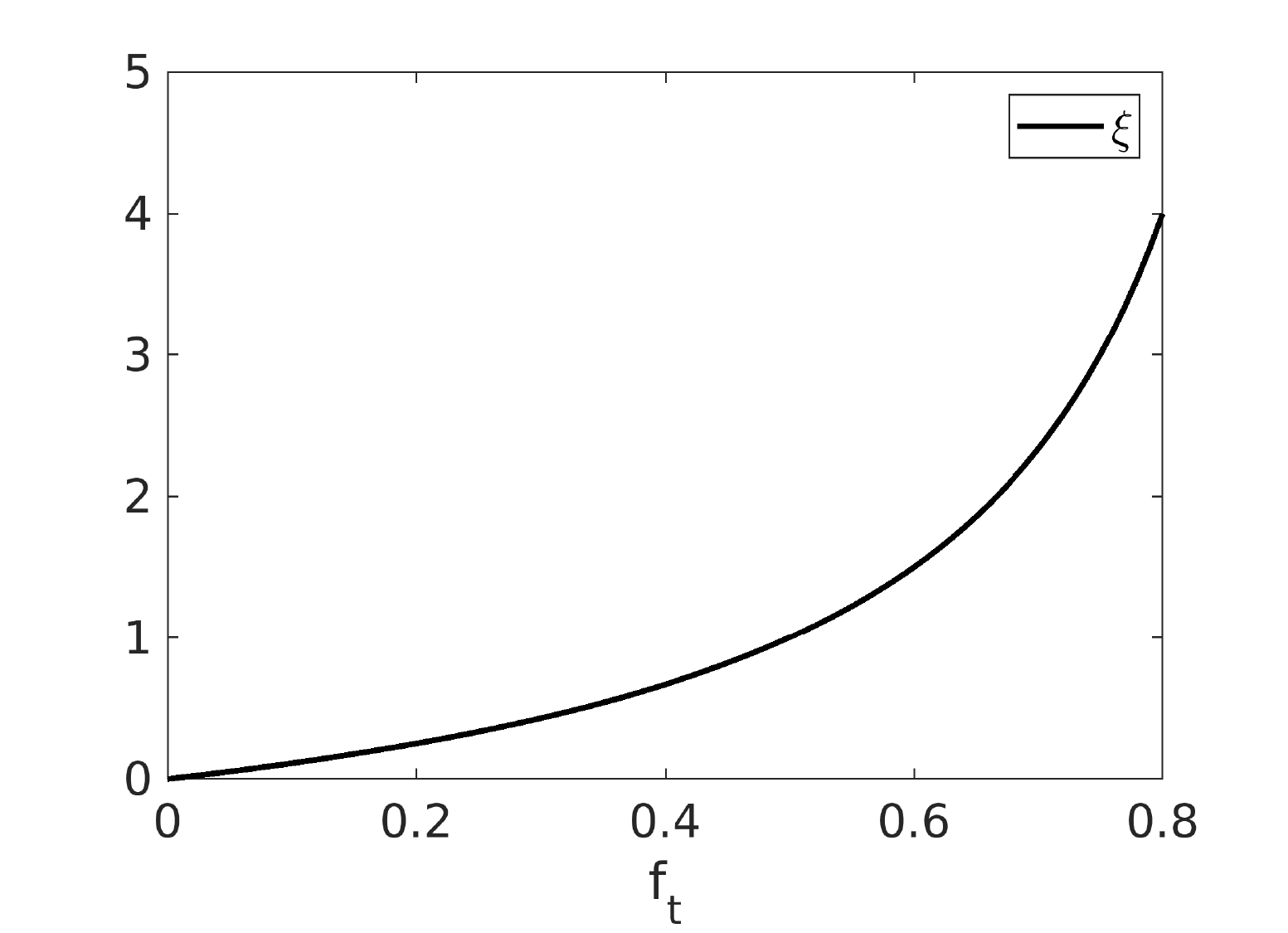}
\caption{Parameter $\ratio = \trap / (1- \trap)$ v.s. trapped-fraction $\trap$.}
\label{fig-ctemratio}
\end{center}
\end{figure}

We will now focus on the `pure' CTEM turbulence, i.e. $\nabla T_e$-driven CTEM. That is, considering quasi-neutrality $n_i = n + (1-\trap) [\phi - \langle \phi \rangle]$, we assume that the mode frequency resonates with the precession-drift frequency, i.e. $\wk \sim \frac{5}{3} \wde$.  Hence, we use the approximation:
\begin{equation}
|n_i| \ll |n|, |\phi - \langle \phi \rangle|
\end{equation}
In this regime, the electron dynamics decouples from the ion dynamics, and the quasi-neutrality condition reduces to \cite{Anderson2006, AndersonNordman2006} :
\begin{equation}
n \simeq - (1- \trap) [\phi - \langle \phi \rangle]
\label{approx2}
\end{equation}
Hence, in this regime, the effective density fluctuations are proportional to the electric potential fluctuations, which implies no turbulent particle transport.
This directly implies that the zonal density vanishes $\dif_t \langle n \rangle = 0$. The dynamics of zonal density was analyzed in Refs. \cite{LeconteKobayashi2021, SinghDiamond2021}.
Using approximation (\ref{approx2}), the system (\ref{den00}-\ref{ti00}) reduces to the following $\nabla T_e$-driven CTEM model:
\begin{eqnarray}
\frac{\dif \tilde \phi}{\dif t} & + & {\bf v}_E . \nabla \tilde \phi - \ratio v_{*e} \frac{\dif \phi}{\dif y} = - \epsn v_{*e} \frac{\dif}{\dif y} [(1+ \ratio) \phi - \ratio \te ]
\label{den1} \qquad\\
\frac{\dif \te}{\dif t} & + & {\bf v}_E . \nabla \te + \etae v_{*e} \frac{\dif \phi}{\dif y} = \frac{2}{3 \ratio} \epsn v_{*e} \frac{\dif}{\dif y} [(1+ \ratio) \phi - \frac{7}{2} \ratio \te ],
\label{temp1}
\end{eqnarray}

with $\tilde \phi = \phi - \langle \phi \rangle$ and $\ratio= \trap / (1- \trap)$. Note that the connection between the slab geometry of the present model and the standard toroidal action-angle variables is explained in Appendix A of Ref. \cite{GarbetDubuitAsp2005}.
The simplified CTEM model (\ref{den1}, \ref{temp1}), which is a limiting case of the original model (\ref{den00}-\ref{ti00}) also conserves energy. 
The ratio $\xi$ is shown v.s. trapped-fraction $\trap$ [Fig.\ref{fig-ctemratio}]. One sees that the parameter $\xi$ is of order unity for $\trap \simeq 0.5$ and increases rapidly for $\trap > 0.5$.

Expressing the potential $\tilde \phi \simeq - n / (1- \trap)$ in terms of effective density $n$ from the quasi-neutrality (\ref{approx2}), and applying a Fourier transform $g = g_k(x,t) e^{i \ky y}$, with $g=\den, \te$, one obtains, after some algebra:
\begin{eqnarray}
\frac{\dif \nk}{\dif t} - i \ratio \dia \nk  & = & - i \wde \Big[ (1+ \ratio) \nk + \tk \Big],
\label{weakturb1} \qquad\\
\frac{\dif \tk}{\dif t} - i \ratio \etae \dia \nk & = & - \frac{2}{3} i \wde \Big[ (1+ \ratio) \nk +  \frac{7}{2} \tk \Big],
\qquad
\label{weakturb2}
\end{eqnarray}
where $\tk =\trap T_{ek}$, and $\wde= \epsn \gele \dia$ is the trapped-electron precession-drift frequency.

\subsection{Schr$\ddot {\bf o}$dinger-like equation for $\nabla T_e$-driven CTEM}

The linear CTEM system (\ref{weakturb1}, \ref{weakturb2}) can be written in the form of a 2-component Schr$\rm \ddot o$dinger-like equation:
\begin{equation}
i \frac{\dif}{\dif t}
\begin{bmatrix}
\nk \\
\tk
\end{bmatrix}
= \hat H
\begin{bmatrix}
\nk \\
\tk
\end{bmatrix},
\label{schrodinger1}
\end{equation}
where the matrix operator $\hat H$ denotes the Hamiltonian operator of the system \cite{ZhouZhuDodin2019}:

\begin{equation}
\hat H =
\begin{bmatrix}
a & b \\
c & d
\end{bmatrix},
\end{equation}
with coefficients:
\begin{eqnarray}
a = \wde - \ratio (\dia - \wde), & & b =  \wde, \\
c =  - \Big[ \ratio \etae \dia - (1+ \ratio) \frac{2}{3} \wde \Big] \le 0, & & d = \frac{7}{3} \wde \qquad
\end{eqnarray}


It is convenient to diagonalize the Hamiltonian matrix. The result is:
\begin{equation}
\begin{bmatrix}
\frac{1}{2} (a+d) +i \sqrt{b |c| + ad - \frac{1}{4} (a+d)^2 } & 0 \\
0 & \frac{1}{2} (a+d) -i \sqrt{b |c| +ad - \frac{1}{4} (a+d)^2 }
\end{bmatrix}
\end{equation}

This corresponds to a transformation to `normal coordinates':
\begin{eqnarray}
\psik^+ & = & \coef \nk^+ + {\tk}^+, \\
\psik^- & = & \coef^* \nk^- + {\tk}^-,
\end{eqnarray}
where the complex-valued parameter $\coef$ is given by:
\begin{eqnarray}
\coef & = & \frac{d-a}{2 |c|} + \frac{i}{|c|} \sqrt{ b |c| +ad - \frac{1}{4} (a+d)^2 }, \nonumber\\
 & = & \frac{ \frac{2}{3} \wde + \frac{\xi}{2} (\dia - \wde) } { \ratio \etae \dia - (1+ \ratio) \frac{2}{3} \wde }
 + i \frac{\gk}{ \ratio \etae \dia - (1+ \ratio) \frac{2}{3} \wde },
\label{def-coef} \qquad
\end{eqnarray}
with $\gk = \dia \sqrt{\ratio \epsn (\etae - \etae^c)}$ the linear growth-rate and $\etae^c$ the linear threshold given by:
\begin{equation}
\etae^c = (1+ \ratio) \frac{2}{3 \ratio} \epsn
- \frac{7}{3 \xi} [ \epsn - \xi (1- \epsn) ]
+ \frac{1}{\xi \epsn} \Big[ \frac{5}{3} \epsn - \frac{\ratio}{2} (1- \epsn) \Big]^2
\label{lin-threshold}
\end{equation}
Note that the normal coordinates can also be obtained following the method of Ref. \cite{SmolyakovDiamondMedvedev2000}.

The Schr$\rm \ddot o$dinger equation (\ref{schrodinger1}) can hence be written in the form:
\begin{equation}
i \frac{\dif}{\dif t}
\begin{bmatrix}
\psik^+ \\
\psik^-
\end{bmatrix}
= H_H
\begin{bmatrix}
\psik^+ \\
\psik^-
\end{bmatrix}
+i H_A
\begin{bmatrix}
\psik^+ \\
\psik^-
\end{bmatrix},
\label{schrodinger2}
\end{equation}
where $H_H = \frac{1}{2} (\hat H+ \hat H^\dagger)$ and $H_A =  \frac{1}{2i} (\hat H- \hat H^\dagger)$ denote the Hermitian part and anti-Hermitian part of the Hamiltonian. They are given, respectively, as:
\begin{eqnarray}
H_H
& = &
\begin{bmatrix}
\frac{5}{3} \wde - \frac{\ratio}{2} (\dia - \wde) & 0 \\
0 & \frac{5}{3} \wde - \frac{\ratio}{2} (\dia - \wde)
\end{bmatrix}, \quad \nonumber\\
H_A
& = &
\gk \sigma_3 =
\begin{bmatrix}
\gk & 0 \\
0 & - \gk
\end{bmatrix},
\label{antiherm1}
\end{eqnarray}

where $\etae^c$ is the linear threshold (\ref{lin-threshold}). Here, $\sigma_3 = [(1,0),(0,-1)]$ is one of the three Pauli matrices \cite{Pauli1926}. The Pauli matrices form a natural base to diagonalize $2 \times 2$ systems, so it is a convenient tool for two-field systems\cite{Dodin-Private}. 

\subsection{Linear analysis}

After some algebra. one obtains the following quadratic linear dispersion relation:
\begin{equation}
\freq^2 - \Big[ \frac{10}{3} \wde - \ratio (\dia - \wde) \Big] \freq
+ \wde \Big[ \ratio \etae \dia + \frac{7}{3} [\wde - \ratio (\dia - \wde)] - (1+ \ratio) \frac{2}{3} \wde \Big] = 0
\label{lin-disprel1}
\end{equation}
The solution for the linearly unstable branch, with $\freq = \wk + i \gk$ is:
\begin{eqnarray}
\wk^{\rm CTEM} & = & \frac{5}{3} \epsn \dia- \frac{\ratio}{2} \dia (1- \epsn),
\label{freq-ctem} \\
\gk^{\rm CTEM} & = & \sqrt{\ratio \etae \dia \wde + \frac{7}{3} \wde [\wde - \ratio (\dia - \wde)]  -(1+\xi)\frac{2}{3}\wde^2  - \wk^2 }  \nonumber\\
& = & \dia \sqrt{ \ratio \epsn (\etae - \etae^c)},
\label{growth-ctem}
\end{eqnarray}
where $\wk = \wk^{\rm CTEM}$ and $\etae^c$ is the linear CTEM threshold, given by Eq. (\ref{lin-threshold}).
For the linearly damped branch, the frequency is $\wk^{\rm CTEM}$ and the linear growth rate is $- \gk^{\rm CTEM} <0$.
The threshold behaviour - a key feature of CTEM - is recovered in the present simplified model. However, the growth-rate and mode frequency are under-estimated, as shown in the following.

\begin{figure}
\begin{center}
\includegraphics[width=0.5\linewidth]{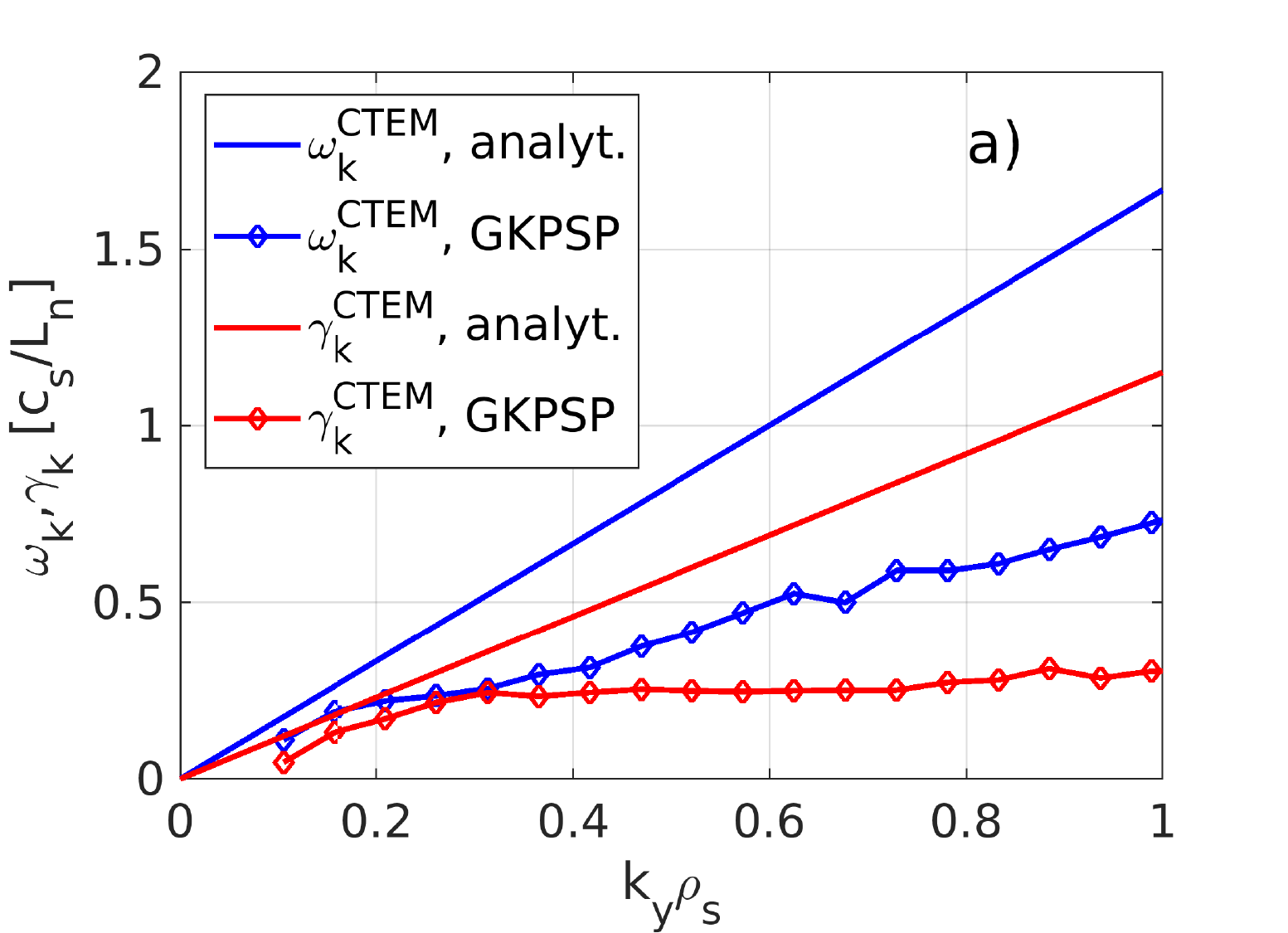}\includegraphics[width=0.5\linewidth]{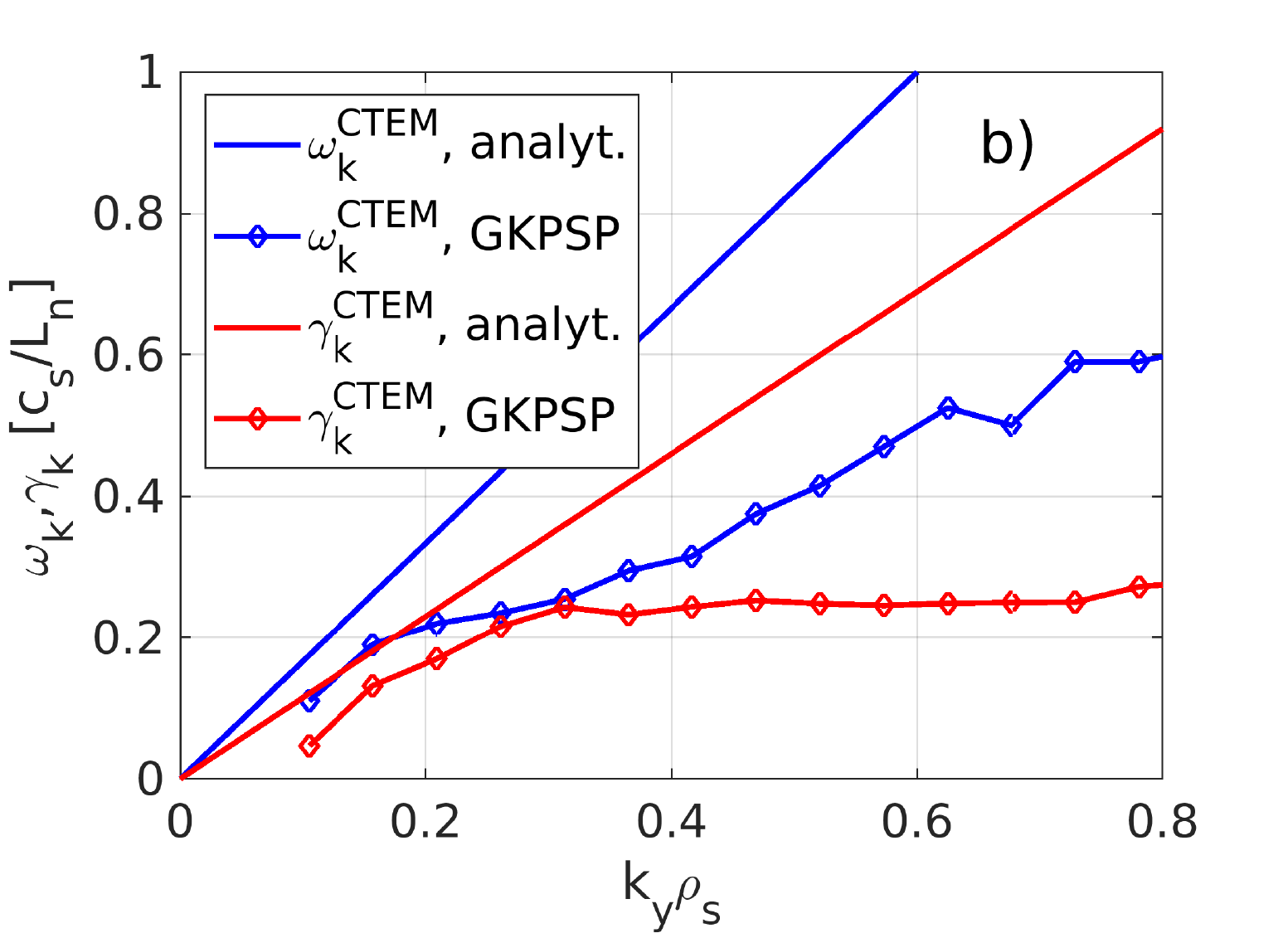}
\caption{a) Linear growth-rate (red) and frequency (blue)  of $\nabla T_e$-driven CTEM two-field model (solid line), and comparison with GKPSP gyrokinetic simulations with bounce-kinetic electrons (diamonds). The parameters are $\etae = 3.1$, $\trap=0.5$ and $\epsn=1$ ($\frac{R}{L_n} =2$), and $\eta_i=1$, and b) close-up for $\ky \rho_s < 0.8$. }
\label{fig-gr-freq-ctem}
\end{center}
\end{figure}

The linear frequency (\ref{freq-ctem}) and growth-rate (\ref{growth-ctem}) are shown for the parameters $\etae=3.1$, $\trap=0.5$ and $\epsn=1$ [Fig. \ref{fig-gr-freq-ctem}], and compared to linear simulations with the global gyrokinetic code GKPSP which includes fully gyrokinetic ions and bounce-averaged kinetic electrons \cite{KwonQiYi2017}. For the GKPSP simulation, the parameters are $\etae=3.1$, $\epsn=1$ ($R/L_n=2$), and $\eta_i=1$. \\
The CTEM frequency (blue) and growth-rate (red) are shown v.s. normalized poloidal wavenumber $\ky\rho_s$ for the fluid CTEM model (square) and GKPSP linear simulations (diamonds).
The fluid CTEM frequency shows only qualitative similarity with the bounce-averaged kinetic result from GKPSP. Namely, both frequencies increase with increasing wavenumber. For low wavenumber $\ky \rho_s < 0.2$, the fluid ITG frequency seems to approximately match the gyrokinetic result.
For the growth-rate also, there is a large discrepancy between the fluid model compared to the GKPSP result, except for $\ky \rho_s < 0.2$. The fluid CTEM model has a growth-rate that increases linearly with $\ky \rho_s$, whereas the growth-rate becomes almost flat at large $\ky \rho_s$ in the GKPSP simulation. The disagreement is not surprising since the fluid $\nabla T_e$-driven CTEM model lacks any inertial/polarization effects ($k_\perp^2 \rho_s^2$). 


\newpage


\subsection{Crossphase dynamics for $\nabla T_e$-driven CTEM}

\begin{figure}
\begin{center}
\includegraphics[width=0.5\linewidth]{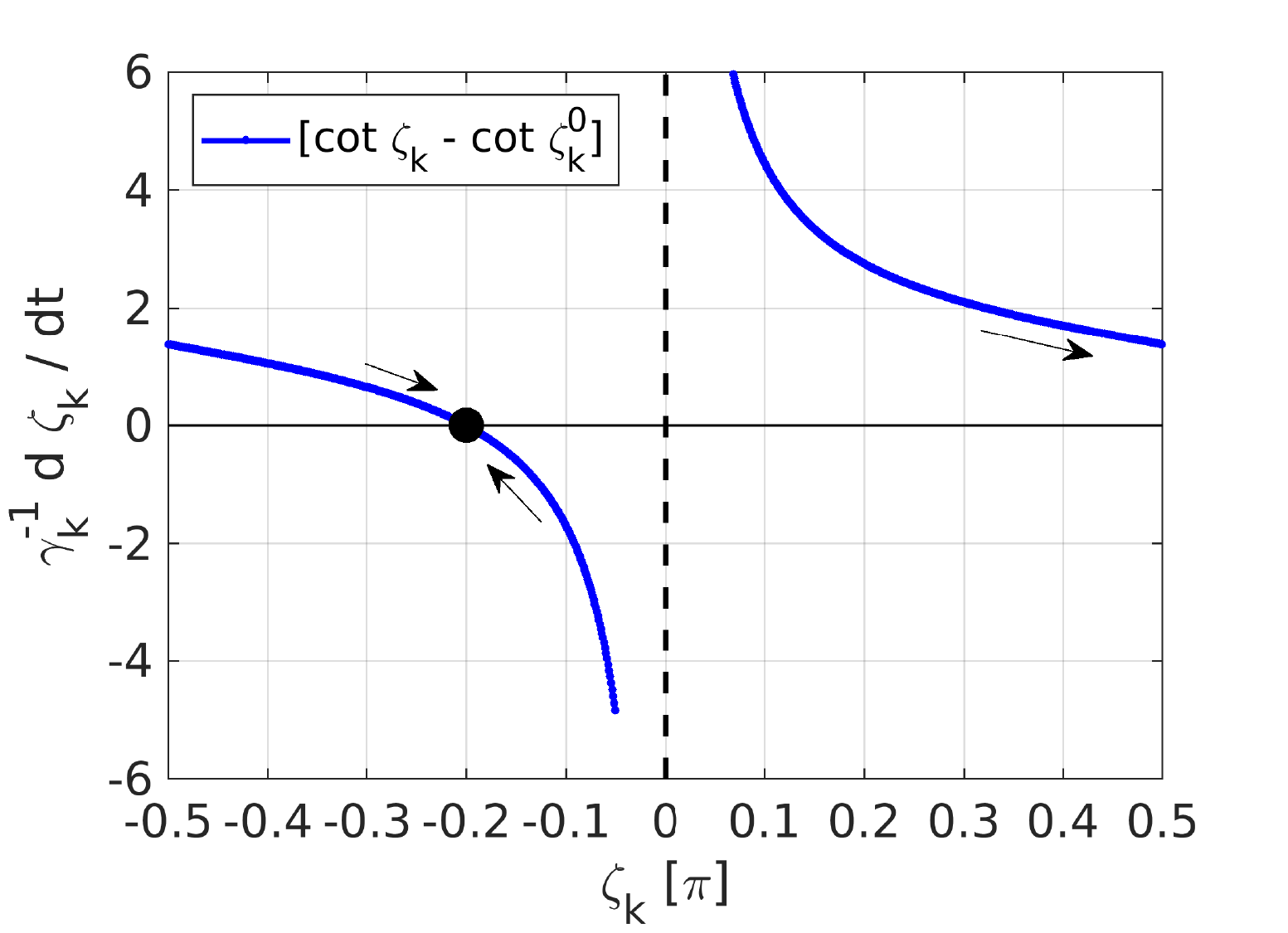}
\caption{Phase response curve $\frac{d \cps}{dt} = f(\cps)$ for the $\nabla T_e$-driven CTEM fluid model. The curve is shown for a value $\cps^0 = -0.2 \pi$ of the phase-locked solution.}
\label{fig-prc-ctem-lin}
\end{center}
\end{figure}

In the $\nabla T_e$-driven CTEM model, there is only a single crossphase $\cps$, the one responsible for electron thermal transport, since particle transport is negligeable in this model. The crossphase between electron temperature fluctuations $\tek$ and potential flustuations $\fk$ is defined as:
\begin{equation}
\cps = \arg (\frac{\fk}{\tek}) = \arg (\frac{\fk}{\sek})
\end{equation}
We now apply to the $\nabla T_e$-driven model (\ref{schrodinger1}) the same analysis as in section 2.3. It is straigthforward to show that the dynamics of the crossphase $\cps$ takes the form:
\begin{equation}
\frac{\dif \cps}{\dif t} =  \wres - \wk +\gk \cot \cps,
\label{cps1}
\end{equation}
where $\cot \cps = 1 / \tan \cps$ is the cotangent of the crossphase, and $\wres = \frac{5}{3} \wde$ is the resonance frequency.
Here, the phase-locked condition $\dif_t \cps =0$ yields the linear crossphase $\cps^0$, used in quasi-linear transport analysis. It is given by:
\begin{eqnarray}
\cps^0 & = & {\rm arctan} \Big( \frac{\gk}{\wk - \wres} \Big) \nonumber \\
 & = & - {\rm arctan} \Big( \frac{ \sqrt{\ratio \epsn (\etae - \etae^c)} }{ \ratio (1- \epsn)} \Big) \le 0,
\label{def-lincps}
\end{eqnarray}
where ${\rm arctan}$ denotes the arctangent function.
Hence the crossphase dynamics for $\nabla T_e$-driven CTEM is similar to the Kuramoto equation for coupled phase-oscillators \cite{Acebron2005}.
In the Kuramoto-like Eq. (\ref{cps1}), the first term on the r.h.s. $\wres - \wk$ is the `entrainment frequency' responsible for \emph{phase-mixing}. Recall that $\wk = \frac{5}{3} \wde - \ratio \dia (1 - \epsn)$ for CTEM. Hence, for CTEM the entrainment frequency takes the form $\wres - \wk = \ratio \dia (1 - \epsn) \ge 0$. The last term on the r.h.s. of Eq. (\ref{cps1}) is the `pinning force', responsible for \emph{phase-locking}. Like in the Kuramoto model, there exists a threshold above which \emph{synchronization} occurs. This synchronization threshold is here given by:
\begin{equation}
\gk \ge |\wk-\wres|
\end{equation}
This seems to implies that, even in the linear theory, a positive linear growth rate $\gk>0$ is only a necessary but \emph{not sufficient} condition for electron heat transport to occur. In addition the linear growth-rate must be larger than the threshold $\gk^{\rm th} =  |\wk-\wres|$. In the fluid CTEM model, this threshold scales like $\gk^{\rm th} \sim 1- \epsn$. This suggests that flat density profiles $\epsn \sim 1$ will lower the synchronization threshold  and trigger phase-locking of the crossphase, and hence electron heat transport, while peaked density profiles $|\epsn| \ll 1$ will require a higher linear growth-rate - higher $\etae$ - to trigger electron heat transport.
In analogy with the `phase reponse curve' associated to oscillators with global couplings, one can define the `crossphase response' curve (CPR) as the r.h.s. of Eq. (\ref{cps1}). The phase response is plotted for a value $\zeta_k^0 = 0.2 \pi$ [Fig.\ref{fig-prc-ctem-lin}]. A similar idea was introduced in Refs. \cite{LeconteSingh2019} and \cite{ZBGuoDiamond2015}.

Note that, for the $\nabla T_e$-driven CTEM model, the turbulent electron heat flux can be written in the form:
\begin{equation}
Q_e = \sum_k \ky \frac{|\fk|^2}{\amptwo} \sin \cps,
\end{equation}
where $\amptwo = |\fk| / |\sek|$ is the amplitude ratio between electron entropy fluctuations $\sek$ and potential fluctuations $\fk$, which takes the form:
\begin{equation}
\amptwo = \frac{\gk}{ (\etae - \frac{2}{3}) \dia \sin \cps}
\label{amp-n3},
\end{equation}

Replacing the amplitude ratio $\amptwo$ by its expression (\ref{amp-n3}), this yields:
\begin{equation}
Q_e = \sum_k \ky \frac{(\etae - \frac{2}{3}) \dia }{\gk} |\fk|^2 \sin^2 \cps
\end{equation}

Remember that $\cps$ is the \emph{instantaneous} crossphase between $\sek$ and $\fk$.
Eq. (\ref{cps1}) can be further linearized around the phase-locked crossphase (\ref{def-lincps}). Using the Taylor series $\cot \cps \simeq \cot \cps^0 + \cot'(\cps^0) [\cps - \cps^0]$, one obtains, after some algebra:
\begin{equation}
\frac{\dif \cps}{\dif t} \simeq - \frac{1}{\tau_k} (\cps - \cps^0), 
\label{cps4}
\end{equation}
where $\tau_k$ is the linear response time (relaxation time) at wavenumber $\bf k$:
\begin{equation}
\tau_k = \Big[ (1+ \cot^2 \cps^0) \gk \Big]^{-1},
\end{equation}
which is of the order of the turbulence correlation time $\tau_c$, i.e. $\tau_k \propto \gk^{-1} \sim \tau_c$, when taking into account resonance-broadening due to turbulence, i.e. $\gk \to \gk + D_t k_\perp^2$, with $D_t$ the turbulent diffusivity . Note that the crossphase dynamics Eq. (\ref{cps4}) has a similar form as the heat flux dynamics of the traffic-jam model of Ref. \cite{Kosuga2014, GurcanDiamondGarbet2013} 
\begin{equation}
\frac{\dif Q}{\dif t} = - \frac{1}{\tau} (Q -Q_0),
\end{equation}
where $Q = \sum_k Q_k$ is the heat flux, $Q_0 = \sum_k Q_k^0$ is the mean heat flux, and $\tau$ is the response time between the instantaneous heat-flux and its relaxed value, which is also of the order of the turbulence correlation time \cite{Kosuga2014, Qi2019}. From the crossphase dynamics Eq. (\ref{cps4}), one sees that the relaxation time becomes very large $\tau_k \to \infty$, when the system is near marginality $\gk \to 0$. Hence, one expects a \emph{slow dynamics} of the crossphase (and associated heat-flux) close to marginality, where the crossphase (and associated heat transport) can remain far from their phase-locked value (the crossphase usually assumed in quasi-linear transport codes), for a significant time. \\
Hence, there seems to be a connection between the `traffic-jam' model of Refs. \cite{Kosuga2014, GurcanDiamondGarbet2013}, and the crossphase dynamics. More work needs to be done to better understand this connection.


\section{Discussion and conclusions}

Let us first discuss the comparison of the fluid model and the linear gyrokinetic simulations using the GKPSP code \cite{KwonQiYi2017}. In the ITG case, the fluid model (\ref{lin-itg11},\ref{lin-itg12}) only shows qualitative similarity with the gyrokinetic simulation. However, the ITG frequency is more closely matched to the gyrokinetic result than the growth-rate, which shows a large difference. Frequency and growth-rate were also compared with Nilsson et al. \cite{Nilsson1990}. The ITG frequency from the ITG model used in this work \cite{Anderson2002} more closely matches the gyrokinetic result compared to that of Ref. \cite{Nilsson1990}, except for $\ky\rho_s \ll 1$.
For the CTEM case also, only qualitative similarity is found between the fluid model and GKPSP simulation, except at low wavenumbers $\ky \rho_s < 0.2$. This is not surprising, since the $\nabla T_e$-driven CTEM fluid model does not contain inertial - i.e. polarization - effects, which are stabilizing at larger wavenumbers $\ky \rho_s \sim 1$. \\
Let us now discuss the Kuramoto-like equation (\ref{cps1}) describing the crossphase dynamics of the fluid CTEM model. Eq.(\ref{cps1}) is very similar to the Kuramoto equation \cite{Acebron2005}, except that the associated phase-response curve is of the form `cotangent' instead of a sinusoid for the Kuramoto model. It thus has period $\pi$ instead of $2 \pi$ for the Kuramoto model. One particular interesting property of the Kuramoto model is the synchronization of coupled oscillators if the coupling is above a certain threshold $K_{th}$ proportional to the entrainement frequency. By analogy, we may say that for CTEM, the transport $crossphase$ - associated to electron heat transport - at different wave-numbers become synchronized when above the threshold. For CTEM, the threshold depends on the difference between the mode frequency and the \emph{resonance} frequency. The form of the Kuramoto-like equation (\ref{cps1}) for the CTEM instability, which is a reactive instability is very different than the one for the collisonal drift-wave instability or the weakly-dissipative trapped electron mode (DTEM) \cite{LeconteSingh2019}. In the latter case, the cotangent function on the r.h.s. of Eq. (\ref{cps1}) is replaced by the (negative of the) tangent function, and it is multiplied by the electron-ion collision frequency $\nu_{ei}$ instead of the linear growth-rate, since $\gk \ll \nu_{ei}$ for collisional instabilities. This may partly explain the difference between the nature of the two types of instabilities. \newline
In conclusion, the nonlinear dynamics of the $\nabla T_e$-driven CTEM model - analyzed linearly here - will be investigated in future work, especially the zonal flows and associated zonal $T_e$ corrugations and their impact on electron heat transport and staircase formation \cite{DifPradalier2015, DifPradalier2017, LeiQi2022, Kosuga2014}.

\section*{Acknowledgements}
The authors would like to thank Jae-Min Kwon, Sumin Yi, Sehoon Ko, P.H. Diamond, I. Dodin and X. Garbet for helpful discussions.
This work was supported by R\&D Program through Korean Institute of Fusion Energy (KFE) funded by the Ministry of Science and ICT of the Republic of Korea (KFE-EN2241-8).



\section*{Appendix: Validity of the $\nabla T_e$-driven CTEM limit}

Linearizing the original system (\ref{den00}-\ref{ti00}) for $\gi = \gele = 1$, one obtains:
\begin{eqnarray}
- i \freq \nk + i \trap \dia \fk = - i \wde ( \nk - \trap \fk + \trap \tek),
\label{ap-lin-den1} \quad\\
-i \freq \tek + i \etae \dia \fk = - \frac{2}{3 \trap} i \wde (\nk - \trap \fk + \frac{7}{2} \trap \tek),
\label{ap-lin-temp1} \qquad\\
-(1- \trap + \kperpsq) i \freq \fk + i (1- \trap - \frac{1+ \etai}{\tau} \kperpsq) \dia \fk = \nonumber\\ i \wde [ (1- \trap)(1 + 1/ \tau)\fk + \tik +(1+ 1/ \tau)\nk + \trap \tek ],
\label{ap-lin-cb1} \quad\\
-i \freq \tik = i \frac{\wde}{\tau} \Big[ \frac{2}{3 \tau} (1 - \trap + \tau) \fk + \frac{2}{3 \tau} \nk + \frac{7}{3} \tik \Big]
- i \Big[ \etai - \frac{2}{3 \tau} (1+ \etai +\tau) \kperpsq \Big] \frac{\dia}{\tau} \fk \nonumber\\
- \frac{2}{3 \tau^2} i \kperpsq \wde \Big[ (1+ \tau) \fk + \frac{\tau}{1- \trap} \tik + \frac{1+ \tau}{1- \trap} \nk + \frac{\tau \trap}{1- \trap} \tek \Big],
\label{ap-lin-ti1} \quad
\end{eqnarray}
with $\dia = \ky v_{*e}$ the electron diamagnetic frequency, $\wde= \epsn \dia$ the precession-drift frequency.
The linear dispersion relation is best obtained by first transforming the system. Adding Eqs. (\ref{ap-lin-den1}) and (\ref{ap-lin-cb1}) yields the ion continuity equation:
\begin{equation}
- i \freq (n_{ik} + \kperpsq \fk ) + \Big( 1 -\frac{1+ \etai}{\tau} \kperpsq \Big) i \dia \fk = i \frac{\wde}{\tau} \Big[ \Big( 1- \trap + \tau \Big) \fk + \nk + \tau \tik \Big],
\label{ap-lin-ion1}
\end{equation}
with, due to quasi-neutrality, $n_{ik} = n_k + (1- \trap) \fk$ the ion density perturbation.
This can also be written as:
\begin{equation}
- i \freq (n_{ik} + \kperpsq \fk) + \Big( 1 -\frac{1+ \etai}{\tau} \kperpsq \Big) i \dia \fk = -i \omega_{di} ( \tau \fk +n_{ik} + \tau \tik ),
\label{ap-lin-ion2}
\end{equation}
with $\omega_{di} = - \wde / \tau <0$ the ion magnetic drift.

The linear ion density response is then:
\begin{equation}
(\freq - \omega_{di}) n_{ik} = \Big[ \dia + \tau \omega_{di} - \kperpsq ( \freq + \ai \dia) \Big] \fk + \tau \omega_{di} \tik,
\label{ap-lin-ion3}
\end{equation}
where $\ai = (1+ \etai) / \tau$ represents ion FLR effects.
Multiplying Eq. (\ref{ap-lin-ti1}) by $3/2$ and substracting Eq. (\ref{ap-lin-ion2}) yields the ion heat balance:
\begin{align}
- i \freq \Big[ \frac{3}{2} \tik -  \frac{1- \trap}{\tau} \fk - \frac{\nk}{\tau} \Big] - \frac{3}{2} i (\etai - \frac{2}{3}) \dia \fk =
- \frac{5}{2} i \omega_{di} \tik
\label{ap-lin-tiresp0}
\end{align}
The associated  linear ion temperature response is:
\begin{equation}
\tik = \frac{1}{\freq - \frac{5}{3} \wdi } \left[ ( \etai - \frac{2}{3}) \dia \fk + \frac{2}{3 \tau} \freq n_{ik} \right]
\label{ap-lin-tiresp1}
\end{equation}

Replacing $\tik$ in terms of $n_{ik}$ and $\fk$ in Eq. (\ref{ap-lin-ion3}), the linear ion density response takes the form:
\begin{equation}
n_{ik} =  \frac{ (\dia + \wdi) (\freq - \frac{5}{3} \wdi ) - \kperpsq ( \freq + \ai \dia) (\freq -\frac{5}{3} \wdi) + (\etai -\frac{7}{3} +\frac{5}{3} \epsn) \dia \wdi }{N_i} \fk,
\label{ap-lin-niresp1}
\end{equation}
with $N_i = \freq^2 - \frac{10}{3} \wdi \freq + \frac{5}{3} \wdi^2$.
Next, we analyze the linear electron dynamics. Multiplying Eq. (\ref{ap-lin-temp1}) by $\frac{3}{2} \trap$ and substracting Eq. (\ref{ap-lin-den1}), one obtains the electron heat balance:
\begin{equation}
- i \freq \left[ \frac{3}{2} \trap \tek - \nk \right] + \frac{3}{2} i \trap \left[ \etae - \frac{2}{3} \right] \dia \fk =
- \frac{5}{2} i \trap \wde \tek
\end{equation}
This yields the linear electron temperature response:
\begin{equation}
\tek = \frac{1}{\trap (\freq - \frac{5}{3} \wde)} \left[ \Big( \etae - \frac{2}{3} \Big) \trap \dia \fk + \frac{2}{3} \freq \nk \right]
\label{ap-lin-teresp1}
\end{equation}
Expressing $\tek$ in terms of $\fk$ and $\nk$, the linear electron density response is then:
\begin{equation}
(\freq - \wde) \nk  = \trap (\dia - \wde) \fk + \frac{\wde}{\freq - \frac{5}{3} \wde} \left[ \frac{2}{3} \freq \nk + \trap \dia \Big( \etae - \frac{2}{3} \Big) \fk \right]
\end{equation}

After some algebra, one obtains:
\begin{equation}
\nk = \trap \Big[ \freq - \wde - \frac{2}{3} \frac{\wde \freq}{\freq - \frac{5}{3} \wde} \Big]^{-1}
\left[ \dia - \wde + \Big( \etae - \frac{2}{3} \Big) \frac{\dia \wde}{\freq - \frac{5}{3} \wde} \right] \fk
\label{ap-lin-denresp1}
\end{equation}
Note the identity $\wde (1 + \frac{5}{3} \delta ) = \freq \delta$, with $\delta = \wde / (\freq - \frac{5}{3} \wde)$ obtained in Ref. \cite{JarmenNordman1992}. The denominator can be rewritten as $\freq - \frac{5}{3} \wde \Big( 1 + \frac{2}{3} \delta \Big) = \freq - \frac{5}{3} (\freq - \wde) \delta$. Then, multiplying both numerator and denominator of expression (\ref{ap-lin-denresp1}) by $\freq - \frac{5}{3} \wde$, one obtains - after some algebra - the following  linear trapped-electron density response:
\begin{equation}
\nk = \trap \frac{ (\dia - \wde) ( \freq - \frac{5}{3} \wde ) + ( \etae - \frac{2}{3} ) \dia \wde }
{ N_e  } \fk,
\label{ap-lin-denresp2}
\end{equation}
with $N_e = \freq^2 - \frac{10}{3} \wde \freq + \frac{5}{3} \wde^2$.
Finally, using quasi-neutrality $n_{ik} = \nk + (1-\trap) \fk$ yields the following linear dispersion relation:
\begin{equation}
\frac{ (\dia + \tau \omega_{di}) (\freq - \frac{5}{3} \omega_{di}) - \kperpsq \freq + (\etai - \frac{2}{3} ) \dia \omega_{di} }{N_i}= \trap \frac{ (\dia - \wde)(\freq - \frac{5}{3} \wde) + ( \etae - \frac{2}{3} )\dia \wde }
{N_e} + 1 - \trap,
\label{ap-lin-disprel1}
\end{equation}
In the limit of pure CTEM, the mode frequency resonates with the precession-drift frequency $\freq \sim \frac{5}{3} \wde$.  In this limit, $N_i \gg N_e$, i.e. $|n_{ik}| \ll (1- \trap)| \fk| ,|\nk|$ and the dispersion relation (\ref{ap-lin-disprel1}) reduces to:
\begin{equation}
\trap \frac{ (\dia - \wde)(\freq - \frac{5}{3} \wde) + ( \etae - \frac{2}{3} )\dia \wde }
{N_e} + 1 - \trap \simeq 0,
\end{equation}

\end{document}